\documentclass{aa} 
\pdfoutput=1

\usepackage{amsmath}
\usepackage[varg]{txfonts}
\usepackage{graphicx}
\usepackage{natbib,twoopt}
\usepackage{xspace}
\usepackage{nicefrac}
\usepackage{aas_macros}
\usepackage{hyperref}  %the draft option supresses colored links temporarily
\usepackage{tablefootnote}
\usepackage{multirow}
\usepackage[usenames, dvipsnames]{color}  % SdM allows for collor comments
\usepackage{placeins}
\usepackage{bm}

\defcitealias{Podsiadlowski+1992}{P92}
\defcitealias{Zapartas+2017}{Z17}

\newcommand{\Msun}{\ensuremath{\,M_{\odot}}\xspace}

\def\pasa{Publ. of the Astronomical Soc. of Australia}                 % Publ. of the Astronomical Soc. of Australia
\def\eprint{}         

\authorrunning{E. Zapartas, S. E. de Mink...}
\titlerunning {Type II}

\begin{document}

\title{The diverse lives of progenitors of hydrogen-rich core-collapse supernovae: the role of binary interaction}

\author{Emmanouil~Zapartas\inst{1,2,*}, Selma~E.~de Mink\inst{1,3}, Stephen~Justham \inst{1,4,5},  Nathan~Smith \inst{6},  Alex~de~Koter \inst{1,7}, Mathieu~Renzo \inst{1}, Iair~Arcavi\inst{8},  Rob~Farmer \inst{1},  Ylva~G\"otberg \inst{1,9} \& 
Silvia~Toonen \inst{1,10}}

\institute{Anton Pannekoek Institute of Astronomy and GRAPPA, University of Amsterdam, 1090 GE Amsterdam, the Netherlands.
    \and Geneva Observatory, University of Geneva, CH-1290 Sauverny, Switzerland.
    %\and GRAPPA, University of Amsterdam, Science Park 904, 1098 XH Amsterdam, The Netherlands.
    \and Center for Astrophysics, Harvard-Smithsonian, 60 Garden Street, Cambridge, MA 02138, USA.
 \and  School of Astronomy and Space Science, University of the Chinese Academy of Sciences, Beijing 100012, China.
 \and National Astronomical Observatories, Chinese Academy of Sciences, Beijing 100012, China.
   \and Steward Observatory, University of Arizona, 933 N. Cherry Avenue, Tucson, AZ 85721, USA.
 \and Institute of Astronomy, KU Leuven, Celestijnenlaan 200 D, B-3001 Leuven, Belgium.
 \and The School of Physics and Astronomy, Tel Aviv University, Tel Aviv 69978, Israel.
 \and The Observatories of the Carnegie Institution for Science, 813 Santa Barbara Street, Pasadena, CA 91101, USA. 
 \and Birmingham Institute for Gravitational Wave Astronomy and School of Physics and Astronomy, University of Birmingham,
Birmingham, B15 2TT, United Kingdom \\
     $^*$\email{ezapartas@gmail.com}
}
  
   \date{The paper has been accepted for publication in Astronomy and Astrophysics.}

\abstract{ 
Hydrogen-rich supernovae, known as Type II (SNe\,II), are the most common class of explosions observed following the collapse of the core of massive stars. 
We use analytical estimates and population synthesis simulations to assess the fraction of SNe\,II progenitors that are expected to have exchanged mass with a companion prior to explosion. 
We estimate that 1/3 to 1/2 of SN\,II progenitors have a history of mass exchange with a binary companion before exploding.  
The dominant binary channels leading to SN\,II progenitors involve the merger of binary stars. Mergers are expected to produce a diversity of SN\,II  progenitor characteristics, depending on the evolutionary timing and properties of the merger. 
Alternatively, SN\,II progenitors from interacting binaries may have accreted mass from their companion, and subsequently been ejected from the binary system after their companion exploded.  
We %repeat our simulations for various assumptions and we find 
show that the overall fraction of SN\,II progenitors that are predicted to have experienced binary interaction is robust against the main physical uncertainties in our models. % to investigate the robustness of our predictions against the main uncertainties. 
However, the relative importance of different binary evolutionary channels is affected by changing physical assumptions. 
We further discuss ways in which binarity might contribute to the observed diversity of SNe\,II  by considering potential observational signatures arising from each binary channel.   
For supernovae which have a substantial H-rich envelope at explosion (i.e., excluding Type IIb SNe), a surviving non-compact companion would typically indicate that the supernova progenitor star was in a wide, non-interacting binary. 
We argue that a significant fraction of even Type II-P SNe are expected to have gained mass from a companion prior to explosion. 
}


\maketitle

\section{Introduction}\label{ch5:sec:intro}

Core-collapse supernovae (SNe) are explosions that occur at the end of the evolution of massive stars and mark the birth of neutron stars and black holes \citep[e.g.,][]{Baade+1934, Bethe+1979, Woosley+2002, Heger+2003}. They can be observed as transients that, in some cases, temporarily outshine the entire host galaxy. We are currently anticipating a wealth of data that should become available from ongoing and near-future automated surveys with robotic telescopes, such as the Zwicky Transient Facility \citep{Bellm2014,Smith+2014a}, the All-Sky Automated Survey for SuperNovae \citep[ASAS-SN,][]{Shappee+2014}, Pan-STARRS \citep{Kaiser+2010}, and the Large Synoptic Survey Telescope \citep[LSST,][]{LSST-Science-Collaboration+2009}. The hope is that these surveys and follow-up campaigns will allow us to address many of the still unanswered questions about these explosions and their progenitors.%One of these questions concerns the origin of the large diversity in the light curves and spectra. 

Observationally, two main groups of core-collapse SNe can be distinguished, ``hydrogen rich supernovae'', that show clear signs of hydrogen in their spectra, and ``stripped-envelope supernovae'', where hydrogen signatures are absent or only present at the early times, see \citet{Filippenko1997} and \citet{Gal-Yam2017} for reviews.  The first group contains a variety of subtypes including II-P (which have light curves that show a distinctive flat plateau), II-L (which show a light curve declining linearly in magnitude), IIn (which show narrow hydrogen lines in their spectra, interpreted as signatures of interaction with circumstellar material) and Type II-peculiar or 87A-like which display a dome-shaped light curve resembling the famous case SN 1987A, see \citet{Arcavi2017} for a recent review. Throughout this paper when we mention ``hydrogen-rich'' or equivalently ``Type II'' supernovae (SNe\,II) we refer to all the subclasses mentioned above. Another group seems to result from progenitor stars that were stripped of most or all of their hydrogen-envelope prior to explosion. This group of ``stripped-envelope'' SN progenitors includes Type IIb (which is a transitional class that only shows evidence for hydrogen at early times), and Type Ib, Ic, Ic-broadlined and Ibn (which show no signatures of hydrogen), see \citet{Pian+2017}.

The progenitors of core-collapse SNe are believed to spend most of their lives as early-B- and O-type stars.  These stars are very often found to be members of close binary systems, as several recent studies have shown \citep{Kobulnicky+2007, Sana+2012, Chini+2012, Kiminki+2012, Sana+2013, Dunstall+2015, Moe+2017, Almeida+2017}. The majority of binary systems are close enough that the progenitor stars are expected to interact with their companion during their lifetime prior to explosion.  This raises the question of how binary interaction affects the final explosion properties and in particular whether binarity plays a central role in the observed diversity among core-collapse SNe \citep[e.g.,][]{Nomoto+1996}.

Binarity is now commonly considered as one of the explanations for stripped-envelope SNe. This is because mass transfer to a companion star provides a natural mechanism for a star to lose its hydrogen-rich envelope \citep[e.g.][]{Kippenhahn+1967,Podsiadlowski+1992,Nomoto+1996, Kobulnicky+2007,Smith+2011,Eldridge+2013}. It provides an alternative to the hypothesis that the progenitors of these hydrogen-poor SNe live in isolation and lose their envelope as a result of stellar winds and eruptive mass loss episodes of very massive (and thus rare) stars \citep[e.g.,][]{Begelman+1986,Gaskell+1986, Georgy+2012, Groh+2013b}. The binary scenario can help explain the high relative rate of stripped-envelope SNe 
\citep[e.g.,][]{Smartt+2009, Smith+2011, Li+2011, Eldridge+2013, Graur+2017a},  
the difficulties to detect their progenitors in pre-explosion images \citep[e.g.,][see however \citealt{Yoon+2012a}, \citealt{Eldridge+2013}, and \citealt{Tramper+2015}]{Van-Dyk+2003,Maund+2005,Maund+2005a,Smartt+2009,Eldridge+2013, Cao+2013, Van-Dyk+2018}, and their low ejecta masses \citep[e.g.,][]{Ensman+1988, Drout+2011, Taddia+2015, Lyman+2016}.  %, Van-Dyk2016

Binarity is less often considered in the case of hydrogen-rich, Type II SNe.  This is probably because binarity, at first glance, does not seem to be needed to explain these events, especially the most abundant Type II-P SNe. 
For example, single stellar models predict stars with initial masses between about 8 and 25\Msun to end their lives as red supergiants \citep[e.g.][]{Woosley+2006, Groh+2013a}. Such stars have extended envelopes that typically contain several solar masses of hydrogen.  Analytical calculations and numerical simulations of the light curves and spectra that would result from the successful explosion of such progenitor stars are able to reproduce the main features reasonably well, including the characteristic plateau in the light curve that is the defining signature of Type II-P SNe \citep{Popov1993,Filippenko1997,Bersten+2011,Dessart+2013,Morozova+2016}. 
  Furthermore, searches for progenitors in pre-explosion images of Type II-P SNe indeed show the presence of presumably single red supergiants at the explosion site in several cases \citep{Van-Dyk+2003,Maund+2005a,Smartt+2009}.

However, the fact that massive stars so often are detected with a close companion implies that many progenitors, even of Type II SNe, will  experience some kind of binary interaction during their life. Indeed, many studies have considered binarity as a possible explanation of unanswered observed characteristics of SNe\,II and  have addressed the role of binarity in the context of hydrogen-rich SNe, from various perspectives, inlcuding \citet[][]{Podsiadlowski+1989,Podsiadlowski+1990,Podsiadlowski1992, De-Donder+2003a, Eldridge+2008, Eldridge+2011, Smartt+2009, Vanbeveren+2013,  Justham+2014, Smith+2014, Zapartas+2017,Soker+2018} and most recently the set of comprehensive simulations by \citet{Eldridge+2018}. 

The aim of this paper is to estimate the importance of the role of binarity in the lives of the progenitors of hydrogen-rich supernovae by (1) identifying the main evolutionary scenarios for single and binary stars that lead to hydrogen-rich SNe, (2) estimating their relative rates,  (3) investigating the robustness of these findings against uncertainties and (4) discussing the possible implications of binarity on the properties of SN\,II progenitors and comparing with the observed rates and statistical properties of these events.

Our paper is organized as follows.  We first present an overview of the main evolutionary channels (Section~\ref{ch5:sec:overview_of_paths}). 
Subsequently, we estimate the relative rates for the different channels using two approaches. Firstly, by performing simple analytical estimates based on idealized assumptions (Section \ref{ch5:sec:analytical_calculation}). Secondly, by comparing with the results obtained from a suite of full binary population synthesis simulations, where we also discuss the robustness of our findings against uncertainties in our model assumptions (Section~\ref{ch5:sec:numerical_simulations}). In the same section we also % and which ones are sensitive  and 
compare our numerical results with our analytical estimates and with the results obtained in earlier theoretical studies. % (Section~\ref{ch5:sec:comparison_with_previous_work}).

We  find that a third to half of Type II SNe progenitors are expected to have experienced mass exchange with a binary companion through Roche-lobe overflow (RLOF), in many cases including the merging of the two stars, before explosion. 
We discuss the possible end fate of the progenitors and the observed SNe\,II for each of these channels, and whether binary evolution can help explain the diversity among SNe\,II (Section~\ref{ch5:sec:comparison_with_observations}).  
We summarize our findings in Section ~\ref{ch5:sec:summary}.

\section{Overview of single and binary evolutionary channels that lead to hydrogen-rich SNe}\label{ch5:sec:overview_of_paths}

Hydrogen-rich SNe can originate from a variety of evolutionary channels.  The simplest scenario concerns the evolution of a massive single star that ends its life as a red supergiant. This scenario has historically received most attention and has been successful in explaining various observed characteristics.  Here we are interested in the additional contribution of evolutionary scenarios that involve interaction with a binary companion.  

Studies that considered binary interaction in the context of core-collapse SNe have focused mainly on the fate of the initially more massive ``primary'' star. This star loses part or all of its hydrogen envelope when it fills its Roche lobe. This channel has therefore been considered as a promising evolutionary path that gives rise to stripped-envelope SNe of Type IIb, Ib and Ic depending on the degree of stripping. If the envelope is not removed entirely, such progenitors could in principle lead to short duration Type II SNe \citep[e.g.,][]{Eldridge+2018}. 

Less attention has typically been given to the fate of the companion star, which is expected to accrete part of the envelope of the primary.   More complex evolutionary channels involve  the merger between the two stars in a binary system. Some mass gaining  companions or the merger products will in principle also give rise to hydrogen-rich supernovae if they retain their hydrogen envelope until the end of its life.

\begin{figure*}[t]\center
  \includegraphics[width=0.75\textwidth]{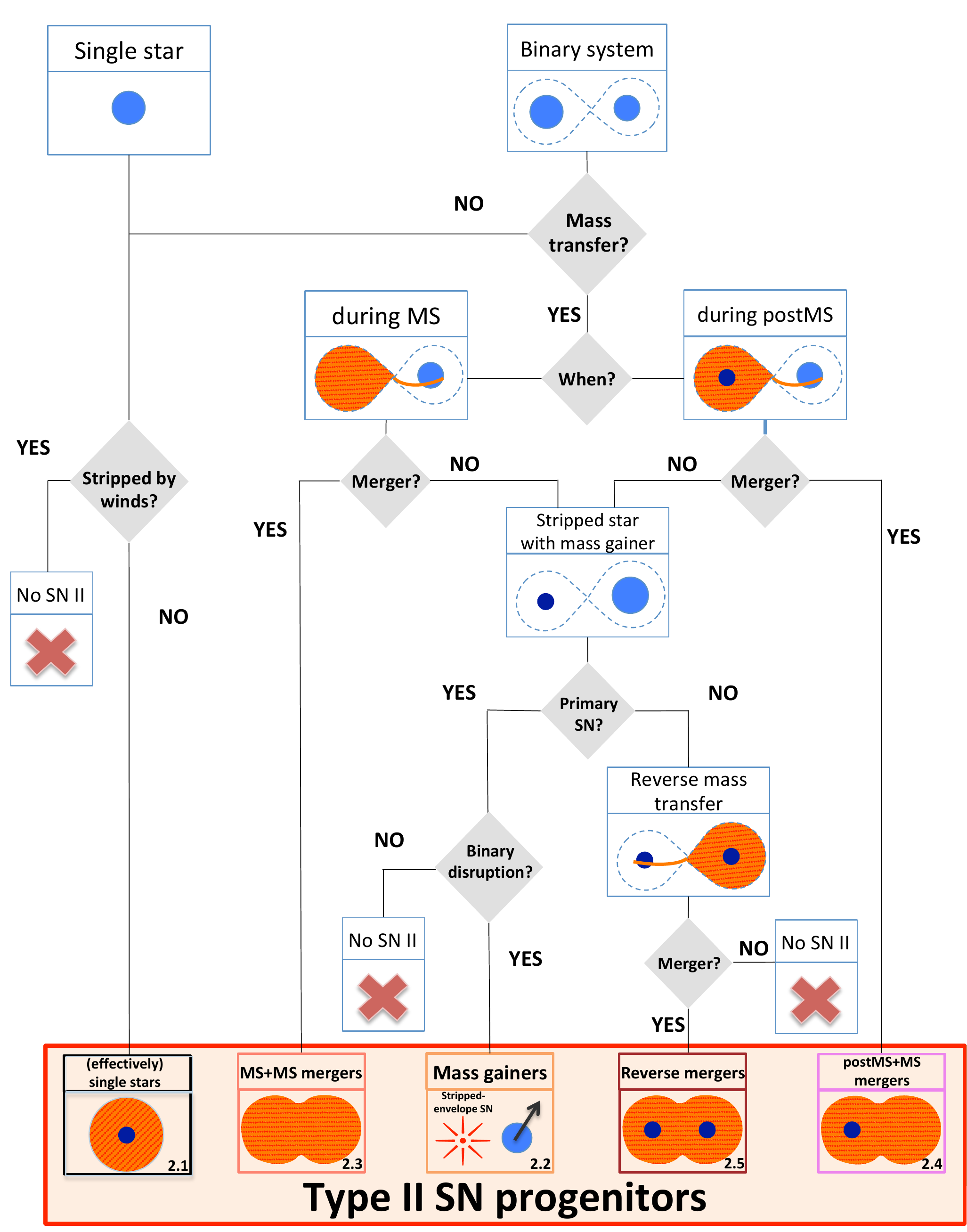}
  \caption{Schematic overview of the five main channels for progenitors of hydrogen-rich core-collapse supernovae (SNe~II):  ``(effectively) single stars'', ``mass gainers'', and three different types of mergers (``MS+MS'', ``postMS+MS'' and ``reverse''). 
  Light blue circles are MS stars, small dark blue circles are hydrogen-deficient cores and orange dashed shapes are hydrogen-rich envelopes, potentially during RLOF or merging.  See the subsections mentioned in the bottom right corner of each path for more details.    \label{ch5:fig:overview_cartoon}}
\end{figure*}

We distinguish five 
main scenarios that give rise to hydrogen-rich SNe, i.e. scenarios that lead to stars with cores massive enough to collapse and that still have  substantial hydrogen-rich envelopes at the moment of explosion. In the remainder of this section we will describe each of these and discuss qualitatively how the hydrogen-rich SNe that result from these channels may differ or be similar to those resulting from single stars. We show a schematic overview in Figure~\ref{ch5:fig:overview_cartoon}. The channels we consider for the progenitors of SNe\,II are:  

\begin{description}
\item[\bf ``Effectively single stars'':] stars that did not interact with a companion either because they are isolated or in wide binary orbits (subsection \ref{sec:1}). 
\item[\bf ``Mass gainers'':] stars that accreted mass from their companion and were then ejected from the binary when the companion exploded (subsection \ref{sec:2}). 
\item[\bf ``Main-sequence mergers:] mergers of two relatively unevolved main-sequence (MS) stars (subsection \ref{sec:3}). 
\item[\bf ``Post-main-sequence mergers:] mergers of a primary star that has evolved off the main sequence (postMS) with a relatively unevolved secondary star (subsection \ref{sec:4}). 
\item[\bf ``Reverse mergers'':] mergers resulting from reverse mass transfer from the evolved secondary with the naked core of the primary (subsection \ref{sec:5}).
\end{description}

Other binary channels may also lead eventually to SNe\,II, for example a donor star that is only partially stripped as a result of stable RLOF \citep{Claeys+2011,Yoon+2017,Gotberg+2017}. Another example involves the initially less massive progenitor spiraling-in but not merging during common-envelope evolution (CEE) triggered by the initially more massive donor star. However, in numerical tests we find these evolutionary scenarios to be rare. 
We discuss them in Section \ref{ch5:sec:numerical_simulations} and list their expected rates in Table \ref{ch5:table:parameters_uncertainties}.

\subsection{Effectively single stars} \label{sec:1}  Single stars with initial masses between about 8 and 25\Msun have historically been considered as the progenitors of hydrogen-rich SNe. The mass boundaries quoted above depend on assumptions in the stellar models concerning internal mixing, mass loss and rotation. Since this channel has been discussed very extensively in the literature, we will be brief here and refer to \citet{Heger+2003} and \citet{Groh+2013a} for discussions.  Stars in very wide binaries are expected to evolve similarly to genuine single stars, and we refer to this extended group of SN\,II progenitors that evolve without binary interaction as ``effectively single stars''.

\subsection{Mass gainers} \label{sec:2}
% (Path A1) 
SN\,II progenitors may originate from stars that have accreted mass through stable RLOF from a companion. These progenitors were the initially less massive secondary star %(the ``secondary" stars) 
of a binary system.  The secondary star is usually still relatively unevolved when the primary fills its Roche lobe, because of its longer evolutionary timescale. The mass gainer is therefore expected to still reside on its main sequence phase in most cases when mass transfer is initiated.

Not all mass gainers are expected to give rise to hydrogen-rich SNe. First, they need to grow a core massive enough to collapse.  Stars that are born with a mass below the threshold for single stars to produce a SN ($M \sim 8\Msun$), can in principle become massive enough after accreting from their companion to eventually give rise to a core-collapse supernova \citep[][hereafter \citetalias{Zapartas+2017}]{Zapartas+2017}. The binary system $\phi$ Persei is thought to be an example of such a system \citep{Schootemeijer+2018}.

The progenitor also needs to retain a large fraction of its hydrogen envelope until explosion. If the stripped remnant of the primary still resides in orbit around the mass gainer, in most cases there will be no space for the mass gainer to expand and become a red supergiant.  Instead, it will fill its Roche lobe and lose its envelope during a phase of reverse mass transfer.  Reverse mass transfer can be prevented if the binary system gets disrupted. This is expected to happen in the large majority of  cases when the primary star dies. The compact object that forms is believed to receive a natal kick, which is typically enough to unbind the system \citep{Eldridge+2011,Renzo+2019}. The secondary star %(which is expected to survive this explosion)
becomes unbound and flies off with a spatial velocity similar to its orbital velocity prior to the explosion \citep{Zwicky1957, Blaauw1961}.  In the minority cases that the primary SN does not unbind the system, the interaction of the secondary with the formed neutron star or black hole will likely remove the envelope and prevent a SN\,II.

Evolutionary calculations show that stars that accrete mass while they are still relatively unevolved are able to readjust their internal structure to their new higher mass. Their convective core grows and new fuel is mixed in to the central burning region. This results in the effective rejuvenation of the accreting star.  After mass transfer ceases and the accretor has restored its thermal equilibrium structure, it is expected to closely resemble the properties of regular but rejuvenated single stars with a mass similar to the new higher mass of the accretor \citep{Hellings1983,Hellings1984, Braun+1995, Dray+2007}. After mass transfer, the accretor typically becomes the more massive star in the system. 

Apart from gaining mass, the accretor is also expected to gain angular momentum and spin-up \citep{Packet1981,de-Mink+2013}. This can result in rotationally induced mixing inside the star, affecting its further evolution \citep[e.g.,][]{Maeder+2000a}. It has been proposed that the rapid rotation rates observed in Be X-ray binary systems is the result of this process \citep[e.g.,][]{Rappaport+1982}. It is uncertain if this enhanced rotation can strengthen a pre-existing magnetic field or generate one. Mass loss through winds may provide a way for the star to lose some of its angular momentum. This may be especially efficient in the presence of a large scale magnetic field, as has been observed for example in the accretor of the interacting Plaskett binary \citep{Grunhut+2013}.  However, in cases where winds are not strong, as for example in low metallicity environments, the star may retain its rapid rotation rate for a much longer part of its evolution. 
If angular momentum can be transported into the core, it is even possible to affect the explosion mechanism, potentially even  resulting in a long Gamma Ray Burst \citep[e.g.,][]{Cantiello+2007}. \citet{Smith+2015} argue that, due to the high rotation, the enrichment and the induced mixing of these stars, they may become Luminous Blue Variables (LBVs) and eventually progenitors of Type IIn SNe.

\subsection{Main-sequence mergers} \label{sec:3}
 An alternative channel toward SNe\,II is the merger of two stars prior to explosion.
We distinguish among different kinds of mergers based on the evolutionary phase of the stars involved in the process. When the two stars merge  during their MS phase, the remnant is expected to evolve as a rejuvenated single star, of higher mass, that continues with central fusion of hydrogen  \citep[e.g.,][]{Glebbeek+2013, de-Mink+2014, Schneider+2016}.  
%similarly to the case of mass accretion onto a secondary star.
The merger process may also lead to high rotational velocities and this can be a cause of enhanced magnetic fields \citep{Schneider+2016}. Thus, its evolution may be similar to the case of mass accretion onto a secondary star, described previously. The degree of rejuvenation depends on the mixing induced by the merger processes.  
%(corresponding to our varied parameter $\mu_{\mathrm{mix}}$, Models 8 and 9 in Table~\ref{ch5:table:parameters_uncertainties}).
A fraction of the total mass is probably lost during the merging process \citep{Lombardi+1995,Lombardi+1996}. %, changing the mass of the newly formed merged star. %(corresponding to our varied parameter of $\mu_{\mathrm{loss}}$, Models 6 and 7) .
Similarly to the previously described scenario, SNe\,II from this channel can originate even from mergers of intermediate mass stars, with initial mass $<M_{\rm min,ccSN}$, if the merger product becomes massive enough \citepalias[e.g.,][]{Zapartas+2017}.

\subsection{Post-main-sequence mergers} \label{sec:4} Hydrogen-rich supernovae may also originate from mergers of a star that has evolved off the MS (postMS) 
%, and is in most cases crossing the Hertzsprung gap (HG), 
with its less massive MS companion. The primary is typically crossing the Hertzsprung gap (HG) but can also be ascending the giant branch. Mergers occur when mass transfer is unstable, leading to a CEE between the two stars. If the envelope is not ejected successfully then the two stars merge. 

As the primary has already formed a compact hydrogen-exhausted core, it is expected to form the center of the merger product due to its higher density.  
The steep chemical gradient at the core-envelope boundary prevents mixing of hydrogen-rich material into the core \citep{Mestel1957,Mestel+1986,Justham+2014}. 
The post main sequence merger product is thus expected to be less rejuvenated than in the previous scenarios, if not at all.

The new star may have an unusual structure. Specifically, the mass of the core relative to that of the envelope may be either larger or smaller than expected for a single star, depending on how much mass was lost during the merger event.  In case of a relatively low core to envelope mass ratio, 
the star may be less extended and thus have a higher surface temperature than an evolved single star of the same core mass. Such mergers may appear as a blue supergiant (BSG) for a significant part of their post-merger evolution. 
It has been suggested that the detected BSG progenitor of the famous SN1987A is a merger product of this type \citep[e.g.,][]{Podsiadlowski+1990, Podsiadlowski1992, Podsiadlowski+1992, Vanbeveren+2013, Glebbeek+2013,Justham+2014, Menon+2017,Urushibata+2018}.

\subsection{Reverse mergers} \label{sec:5}
Mergers can also be triggered by unstable reverse mass transfer of the secondary towards the primary. 
In most cases 
the secondary has already gained mass during one or more prior mass transfer phases from the primary, which has lost its hydrogen-rich envelope  in the process and has become a stripped He core or a white dwarf  at the moment of merger. Therefore, both stars evolved in the process have already developed a hydrogen-exhausted core. 

The end fate of this merger product is highly uncertain but it is anticipated to be an evolved massive star, very close to the end of its life, possibly leading to core collapse \citep[][\citetalias{Zapartas+2017}]{Sparks+1974,De-Donder+2003a,Sabach+2014}. 
We only expect this type of merger in systems where the primary did not undergo core collapse to form a neutron star or black hole.  Mergers in the latter systems may potentially result in the formation of Thorne-$\dot{\mathrm{Z}}$ytkow objects \citep{Thorne+1977}.

 \section{Analytical estimate of the relative rates}
 \label{ch5:sec:analytical_calculation}

Before discussing the outcomes of our numerical population synthesis simulations, we present the results of a simplified analytical estimate of the expected relative contribution of each evolutionary channel discussed above.
These simple estimates cannot replace the results from a more sophisticated simulation that takes into account many different aspects of single star and binary evolution. However, they can provide insight into which channels are likely the main contributors to each scenario without relying on the elaborate set of assumptions that enter into the detailed simulations. 

The fate of a binary system is mainly characterized by three initial parameters: the mass of the primary star, $M_1$, the mass ratio between the two stars, $q = M_2/M_1$, and the orbital period of the system, $P$. Other physical parameters, such as the eccentricity or the spins of the stars are expected to play only a secondary role \citep[e.g.,][]{de-Mink+2015} and for simplicity we do not consider them here.  
Below we estimate the rate of SN\,II progenitors that followed a specific evolutionary channel, based on simple assumptions about the initial parameters of binary systems that lead to each channel. 

\begin{figure*}[t]\center
\includegraphics[width=0.80\textwidth]{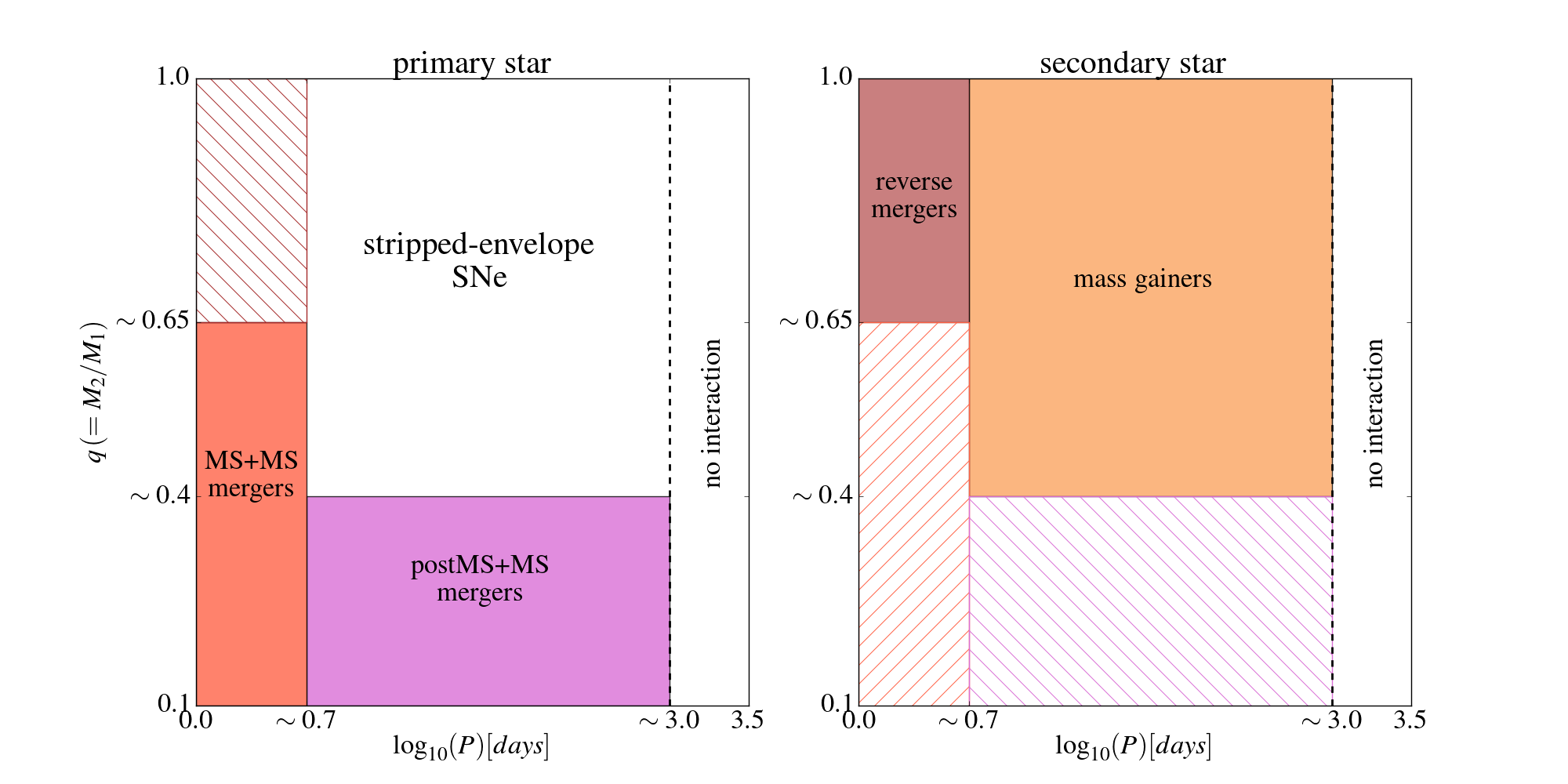}
\caption{Graphical representation of the constraints on the initial parameters that we have assumed for our simple analytical estimate discussed in section \ref{ch5:sec:analytical_calculation}. The figure shows the projection of the initial binary parameter space in the period ($\log_{10}(P)$) - mass ratio ($q=M_2/M_1$) plane. The left/right panel represents the final outcome of the primary/secondary star, or the merger star that was formed after mass transfer from that component. The colored boxes show the simplified limits of the area from which each evolutionary channel originates. The hashed regions show the part of the parameter space that leads to merging triggered by the companion star in each case. 
}
\label{ch5:fig:parameter_space_cartoon}
\end{figure*}

Figure \ref{ch5:fig:parameter_space_cartoon} shows the projection of the birth binary parameter on the initial orbital period and mass ratio ($P-q$) plane for each evolutionary path that is discussed here. 
We focus on these two parameters because they are the most important in determining the evolution of a binary system. For a given mass ratio, the orbital period dictates to a great extent how large in radius the primary donor star needs to be to fill its Roche lobe.  Stars in wider orbits need to expand more to initiate RLOF, which in general implies that they will be in a more advanced phase in their evolution at that moment.  
In addition to the evolutionary phase of the donor, the outcome of binary interaction depends also on the response of the companion star and on the impact of mass exchange on the orbit. For both effects, mass ratio plays a crucial role \citep[e.g.,][]{Schneider+2015}.

We assume here that the initial distributions of $M_1$, $q$, and $P$ are independent of each other, as it had been treated in most studies until now \citep[e.g.,][see however \citealt{Moe+2017}]{Sana+2012,Duchene+2013}.  
The fraction $X_i$ of SN\,II progenitors that followed a binary evolutionary path $i$ will therefore be proportional to the product of the integrals of the initial distributions of these three parameters that this channel is expected to originate from. 
$X_i$ is also by definition proportional to the initial frequency of binary systems, usually referred to as binary fraction, $f_{\rm bin}$, (such that $1-f_{\rm bin}$ is the fraction of systems that consist of a single star at birth). 
To estimate the relative rate of each path, we choose the rate of single stars as the reference point. 
Thus, we approximate the fraction $X_i$ of SN\,II progenitors that followed the binary evolutionary path $i$ as: 
 
\begin{equation}\label{ch5:eq:fraction_prop2}
 X_i = C \times f_{M_1, i}  \times f_{q, i} \times f_{P, i} \times  f_{\rm bin}.  
\end{equation}

The normalization factor, $C$, is chosen such that the sum of all the fractions, $\sum\nolimits X_{\rm  i}$, equals unity. For the assumptions that we make in this section, after estimating each channel, we find that $C = 1.12$.

The probability that a binary with a primary of initial mass in the range $[M_1 , M_1+dM_1]$ is formed follows the initial mass function (IMF), which for this calculation is assumed to have a slope of $\alpha_{\rm IMF} = -2.3$ \citep{Kroupa2001}. SN\,II progenitors that follow single stellar evolution paths originate from initial masses of $8 \lesssim M_1/\Msun \lesssim 25$ \citep[e.g,][]{Heger+2003}. %,Smartt+2009
Thus, 
\begin{equation}
 f_{M_{1},i} = \mathrm{IMF}_{M_{1,i\mathrm{,min}}}^{M_{1,i\rm,max}} = 
  \frac{\int_{M_{1,i\rm,min}}^{M_{1,i\rm,max}}M_1^{-2.3}\,\mathrm{d}M_1}{\int_{8}^{25}M_1^{-2.3}\mathrm{d}M_1}.
\end{equation}
%\frac{\int_{M_{1\rm,i,min}}^{M_{1\rm,i,max}}\mathrm{IMF}\,\mathrm{d}M_1}{\int_{M_{1\rm,single,min}}^{M_{1\rm,single,max}}\mathrm{IMF\,}\mathrm{d}M_1} =

The mass ratio $q$ follows an initial distribution, IQF, which we assume to be flat for the full range of mass ratios considered  of $0.1 < q=M_2/M_1 < 1$ \citep[e.g.,][]{Sana+2012}. 
Thus,

\begin{equation}
 f_{q ,i} = \mathrm{IQF}_{q_{i\mathrm {,min}}}^{q_{i\mathrm {,max}}} = 
 \frac{\int_{q_{i\mathrm {,min}}}^{q_{i\mathrm {,max}}}\,\mathrm{d}q}{\int_{0.1}^{1}\,\mathrm{d}q}.
\end{equation}

The initial period distribution, IPF, is assumed to be a flat distribution in logarithmic space \citep[e.g.,][]{Kiminki+2012} for the full range of periods considered ($0 < \log_{10} (P/\mathrm{day}) < 3.5$, following the boundaries of \citealt{Sana+2012}). So, 

%\begin{align}
\begin{equation}
 f_{P ,i} = \mathrm{IPF}_{\log_{10} P_{i\mathrm{,min}}}^{\log_{10} P_{i\mathrm{,max}}} = 
 \frac{\int_{\log_{10} P_{i\mathrm{,min}}}^{\log_{10} P_{i\mathrm{,max}}}\,(\mathrm{d}\log_{10} P)}{\int_{0}^{3.5}\,(\mathrm{d}\log_{10}P)} .
\end{equation}

For this estimate we adopt a constant, mass-independent binary fraction of $0.5$ for the period limits considered ($0 < \log_{10} (P/\mathrm{day}) < 3.5$), which may even be on the conservative side for the mass ranges of SN progenitors that we focus on \citep[e.g.,][]{Sana+2012,Duchene+2013}. 
Even if the binary fraction is higher when we include wider orbits, these wide systems will not lead to interaction, so they are  considered as effectively single stars here for the purpose of their evolution.

Using Eq.\,\ref{ch5:eq:fraction_prop2}, we calculate below the estimated fraction of each channel discussed in section \ref{ch5:sec:overview_of_paths}.

\paragraph{Single stars  ---} 
For single stars, $f_{M_1 \rm, i}$ is 1 by definition and $f_{q \rm, i} = f_{P \rm, i} = 1$ because no companion is present for this channel. 
As $f_{\mathrm{bin}}=0.5$, then $1-f_{\mathrm{bin}}=0.5$  too, which is used for this channel. So, $X_{\rm single \, stars} = C \times 0.5 \sim 56\%$.

Note that in this calculation, for reasons of simplicity, we do not take into account possible progenitors that are in very wide binary systems and do not exchange mass with their companions \citep[expected for periods of roughly $3.0 \lesssim \log P \lesssim 3.5$;][]{Claeys+2011,Yoon+2017}. These stars would also evolve as ``effectively single stars" and would contribute to this path involving no binary interaction. They are taken into account in our population synthesis findings in section \ref{ch5:sec:fractions}. %49%

\paragraph{Main-sequence mergers  ---} 
Mergers of two main-sequence stars are expected to originate from binary systems of short orbital periods typically of a few days    \citep[roughly $0 \lesssim \log P \lesssim 0.7$;][]{Pols1994,Nelson+2001,de-Mink+2007} in order for the primary to start mass transfer before completing central hydrogen burning. Therefore, $f_{P \rm, i} = {\rm IPF}_0^{0.7}  \sim 0.2$. Merging is expected to take place in binary systems of unequal mass stars, with $q \lesssim 0.65$ for this period range \citep{de-Mink+2007}, thus $f_{q \rm, i} \sim {\rm IQF}_{0.1}^{0.65} \sim 0.61$. %For visualization we show the projection in the $P-q$ plane of the simplified idea of the initial binary parameter space that we discuss in this section in the two panel of Figure \ref{ch5:fig:parameter_space_cartoon}.

The initial binary parameter space that leads to SNe\,II from main-sequence mergers is depicted in orange in the left panel. %, which represents the final outcome of the primary star (or of a merger that was formed when the primary initiates mass transfer, as in this case).
Since SNe\,II can originate from mergers involving intermediate mass stars in case the merger products become massive enough to explode \citepalias[e.g.,][]{Zapartas+2017}, we can make the rough assumption that $f_{M_1 \rm, i} \sim {\rm IMF}_7^{25}  \sim 1.25$, to allow for a contribution of intermediate mass stars. We negelect here the posibility that some stars close to the upper mass threshold for Type II SNe can become Wolf-Rayet stars instead due to the increase in mass of the merger product. %\sim {\rm IMF}_7^{25}
Taking into account also that $f_{\mathrm{bin}}=0.5$,  from Eq.\,\ref{ch5:eq:fraction_prop2} we get that $X_{\rm MS+MS \, mergers} \sim C \times 0.076 \sim 8.5\%$. %7%
%\MZ{maybe I can also depict the parameter space of $M_1$}
%

\paragraph{Post-main-sequence mergers ---} 
Mergers of evolved stars with their MS companions originate from wider systems than main-sequence mergers ($\log P \gtrsim 0.7$). We thus roughly estimate $f_{P \rm, i} \sim {\rm IPF}_{0.7}^{3.0} \sim 0.66$, using $\log_{10}P=3$ as the maximum initial period for binary interaction \citep[e.g.,][]{Claeys+2011,Yoon+2017}.  
For merging to occur, a CEE phase needs to be initiated.  Extreme mass ratio systems are prone to unstable mass transfer and thus to CEE, and although the exact value of the boundary is not well-constrained, we assume that this occurs for evolved donors in systems with $q \lesssim 0.4$, following previous works such as \citet{Wellstein+2001}, \citet{Hurley+2002} and \citet{de-Mink+2013}. In principle CEE can alternatively lead to the ejection of the envelope, but in this simple estimate we assume that all these systems eventually merge, leaving some hydrogen-rich layers on the surface of the formed star. This is consistent in most of the cases with the findings from our computational simulations for our standard assumption of using the entire orbital energy change to eject the envelope ($\alpha_{\rm CEE}=1$, as we will introduce and discuss in section \ref{ch5:sec:model}). Thus, $f_{q \rm, i} \sim {\rm IQF}_{0.1}^{0.4} \sim 0.33$. The projection of the initial parameter space for this channel at the $\log P - q$ plane is depicted in magenta on the left panel of Figure \ref{ch5:fig:parameter_space_cartoon}.   The donor star in such systems originate from a roughly similar part in $M_1$ space as main-sequence mergers ($f_{M_1 \rm, i} \sim {\rm IMF}_7^{25} \sim 1.25$). In contrast, binaries of less extreme mass ratio are able to either follow a phase of stable mass transfer onto the secondary star or are assumed to always survive a CEE by ejecting the envelope, avoiding a merger. Such donors are expected to eventually produce a hydrogen-poor, stripped-envelope SN, not contributing to the SN\,II population that we focus on in this study. 
 We find $X_{\rm postMS+MS \, mergers} \sim  C \times 0.136 \sim 15\%$. %14%

\paragraph{Mass gainers  ---} 
SN\,II progenitors that gained mass through accretion are produced from systems in which an earlier SN occurred from the primary, probably of hydrogen-poor type. In addition, that prior SN needs to disrupt the system before the SN\,II from the secondary, otherwise the system will follow a reverse mass transfer phase. According to \citet{de-Donder+1997}, \citet{Eldridge+2011} and \citet{Renzo+2019}, disruption of the system occurs in roughly 80\% of the cases after SN from the primary star. Binary disruption is significantly more probable for wider systems at the moment of explosion, as the gravitational potential energy needed to be succumbed is lower for larger  separation. Although the binary separation at the moment of explosion is not similar to the initial one, systems formed in wide orbits are expected to end up in even wider orbits in general after a stable phase of mass transfer. %(although CEE from wide systems of $\log_{10}P$ close to $3.0$ follow the opposite trend). 
For the purpose of our estimate we assume that all systems with initial period $0.7 \lesssim \log_{10} P \lesssim 3.0$ will get disrupted whereas the rest will not. So, $f_{P \rm, i} \sim {\rm IPF}_{0.7}^{3.0} \sim 0.66$. 
Also, areas of origin of the previously discussed merger paths are excluded. 
Thus, we consider mass ratios $q \gtrsim 0.4$ and $f_{q \rm, i} \sim {\rm IQF}_{0.4}^{1.0} \sim  0.66$. The initial parameter space for this path is depicted in light brown in the right panel of Figure \ref{ch5:fig:parameter_space_cartoon}, as it is the secondary star that leads to a SN\,II.
If the secondary is initially below the mass threshold of $8 \Msun$ for core-collapse,  it needs to accrete part of the primary's mass to become massive enough to explode as well. This is more likely for higher mass primaries because they can potentially transfer more mass onto their companions, which are on average more massive initially anyway. 
Thus, we roughly assume that this path requires $M_1 \gtrsim 10 \Msun$, thus $f_{M_1 \rm, i} \sim {\rm IMF}_{10}^{25} \sim 0.67$, which results in $X_{\rm mass \, gainers} \sim  C \times 0.146 \sim 16\%$.

\paragraph{Reverse mergers  ---}
Finally, reverse mergers mostly originate from intermediate mass systems in which the primary does not explode before the merger but instead leaves a stripped core or even a white dwarf \citepalias{Zapartas+2017}. Consequently, the primary mass $M_1$ is below the mass threshold of $8\Msun$ for SNe or slightly above if it loses mass during its MS, thus not forming a massive enough core to collapse. We assume here that they originate from binary systems with $5<M_1/\Msun<9$, which leads to $f_{M_1 \rm, i} \sim {\rm IMF}_5^9  \sim 1.27$.
 As the secondary in these systems initially has an even lower mass, it needs to accrete a considerable amount of gas during the first mass transfer phase from the primary, in order to surpass the SN mass threshold. This can be achieved mainly in close binary systems where mass transfer occurs at the early evolutionary stages of the stars and is expected to be more conservative. Thus, here we assume that they originate predominantly from systems that interact the first time during the primary's MS, with $\log_{10}P \lesssim 0.7$. At the same time, early merging during  the initial mass transfer from the primary needs to be avoided, so we exclude the appropriate region of the parameter space, allowing only for $q \gtrsim 0.65$, as shown with dark  brown in the right panel of Figure \ref{ch5:fig:parameter_space_cartoon}. This results in $f_{P \rm, i} \sim  {\rm IPF}_0^{0.7} \sim 0.2$ and $f_{q \rm, i} \sim  {\rm IQF}_{0.65}^{1.0} \sim 0.27$.
So, $X_{\rm reverse \, mergers} \sim C \times  0.035 \sim 4\%$. %14%

\vspace{1em}

In summary, based on these simple estimates for a population including 50\% binary systems, we estimate the probability of a SN\,II progenitor to experience a certain evolutionary path to be as follows:

\begin{equation}
\begin{array}{@{}ll@{}}
 1) &  \bm{X_{\rm single\, stars}\sim 56\%}, \\
 \end{array}
 \end{equation}
 and  
 \begin{equation}
   \left.
  \begin{array}{@{}ll@{}}
 2) & \bm{X_{\rm mass \, gainers}\sim\,16\%}, \\ 
 3) & \bm{X_{\rm MS+MS \, mergers} \sim 8.5\%}, \\
 4) & \bm{X_{\rm postMS+MS \, mergers} \, \sim 15\%}, \\
 5) & \bm{X_{\rm reverse \, mergers} \sim 4\%}.
 \end{array} \right\} \bm{X_{\rm binary \, paths} \sim 44\%}
\end{equation}

This indicates that, while single stellar evolution scenarios are common, the fraction of SN\,II progenitors that are expected to follow binary channels is very significant, possibly about half. 
One important reason for this is that, according to our simple estimates, most massive binaries are expected to lead to \emph{one} Type II SN event, either from the mass gaining star or from a hydrogen-rich merger.  

\paragraph{Stripped-envelope SNe  ---}
In this work  we do not focus on stripped-envelope SNe, but we can use the same simple framework to estimate their relative rate compared to SNe\,II. This ratio can directly be compared to the findings of transient surveys after correcting for biases. 

Single stars of mass $M_1 \gtrsim 25 \Msun$ get stripped due to wind mass loss, contributing to stripped-envelope SNe \citep[although the exact mass threshold is model-dependent, e.g.,][]{Heger+2003,Georgy+2012}. We assume that binaries with donors in the same mass range will also lead to one stripped-envelope event. This is justified  because in almost all cases there will be a binary product (the mass loser, the mass gainer or the merger product) that will not retain its hydrogen-rich envelope up to explosion because of its high wind mass loss rate. 
For simplicity we neglect the possibility of both stars in the binary to end up in stripped-envelope SNe. Following the IMF we get that $f_{M_1 \rm, i} \sim {\rm IMF}_{25}^{100} \sim 0.25$, where we adopted the typical the upper limit of  100 \Msun for stellar populations without significantly affecting the value.

The main binary path for stripped-envelope SNe involves the removal of the hydrogen-rich layers of the primary due to stable RLOF onto the secondary companion. This evolutionary scenario is similar to that followed by systems that produce the mass gainer progenitors for SNe\,II, because it is the explosion of the stripped-envelope primary that disrupts the binary. We use the same assumptions for the boundaries in the parameters space of $P$ and $q$ as in that channel, having $f_{P \rm, i} \sim {\rm IPF}_{0.7}^{3.0} \sim 0.66$ and $f_{q \rm, i} \sim {\rm IQF}_{0.4}^{1.0} \sim  0.66$. Stripping can also occur during a CEE that avoids merging (favored in initially wide orbits), but this should be a minor contribution which we will neglect for this simple estimate. The progenitor needs to be massive enough at birth to explode eventually, with $M_1 > 8 \Msun$ and thus $f_{M_1 \rm, i} \sim {\rm IMF}_{8}^{25} = 1$.

Thus combining our estimates for these two paths toward stripped-envelope SNe  we get that the ratio of them to SNe\,II:

\begin{align}
 \pmb{ R_{\rm stripped / II}} &= X_{\rm stripped} / X_{\rm SNII, \, all \, paths}  = X_{\rm stripped} / 1 =  \nonumber \\
  = C &\times [({\rm IMF}_{8}^{25} \times {\rm IQF}_{0.4}^{1.0} \times{\rm IPF}_{0.7}^{3.0} \times f_{\rm bin}) + {\rm IMF}_{25}^{100}] \sim  \pmb{0.52	}
\end{align}

So SNe~II are expected to occur roughly 2 times more than stripped-envelope ones according to our analytical estimate. 
%So SNe\,II are expected to account for roughly two third of all core-collapse SNe.
This is consistent with observations as we discuss in Section \ref{ch5:sec:comparison_with_observations}.

\section{Results from population synthesis simulations}\label{ch5:sec:numerical_simulations}

In this section we present the results from our numerical population synthesis simulations that explore the significance of the different evolutionary routes toward Type II SN progenitors. 
We describe our computational method in Section \ref{ch5:sec:model}. We present and discuss our findings and their robustness in Section \ref{ch5:sec:fractions}. We compare our numerical results with our analytical estimate in Section \ref{ch5:sec:comparison_with_analytics} and with previous theoretical studies in Section \ref{ch5:sec:comparison_with_previous_work}.

\subsection{Computational method}\label{ch5:sec:model}

We use the stellar population synthesis code {\tt binary\_c}, developed by \citet{Izzard+2009,Izzard+2006,Izzard+2004}, including updates by \citet{de-Mink+2013} and \citet{Schneider+2015},  as well as  \citetalias{Zapartas+2017} and \citet{Zapartas+2017a} that focus on core-collapse SNe.  The code allows us to simulate the evolution of stars based on analytical fitting formulae \citep{Hurley+2000} to the detailed single stellar models by \citet{Pols+1998}, accounting for possible binary interactions \citep{Hurley+2002}.

With this code we simulate the evolution of populations of both single and binary stellar systems. 
We follow the evolution of each star from the zero-age main sequence (ZAMS) to the point that it forms either a white dwarf or explodes as a SN, leaving behind a neutron star or black hole. In this study we  focus on the SN progenitors that have retained a hydrogen envelope at the moment of explosion and will produce a SN\,II. The rest of the core-collapse events originating from progenitors without a hydrogen-rich envelope are counted as stripped-envelope SNe.

We follow a similar approach as in \citetalias{Zapartas+2017}, 
adopting the same treatment of physical processes and the same distributions for the initial parameters of single and binary systems. 
We account for the uncertainties in our input assumptions by varying one at a time the parametrized physical processes and initial conditions of the population. Below we briefly describe the adopted standard assumptions as well as the variations that are most relevant for this work. We refer to \citetalias{Zapartas+2017} and references therein for a more extensive description and discussion of the simulations.  A list of all the variations in our input assumptions can be found in Table~\ref{ch5:table:parameters_uncertainties}, where we also link each varied parameter with the corresponding simulation of \citetalias{Zapartas+2017} whenever possible.

We account for mass loss due to stellar winds in different stages of stellar evolution \citep{Vink+2000,Vink+2001,Nieuwenhuijzen+1990,Hamann+1995} as described in \citetalias{Zapartas+2017}. We treat the uncertainty in the winds \citep[][for a review]{Smith2014} by multiplying the mass loss rate by a universal factor $\eta = 0.1, 0.3, 3$ in our model variations.

\begin{figure*}[t]\center
\includegraphics[width=0.68\textwidth]{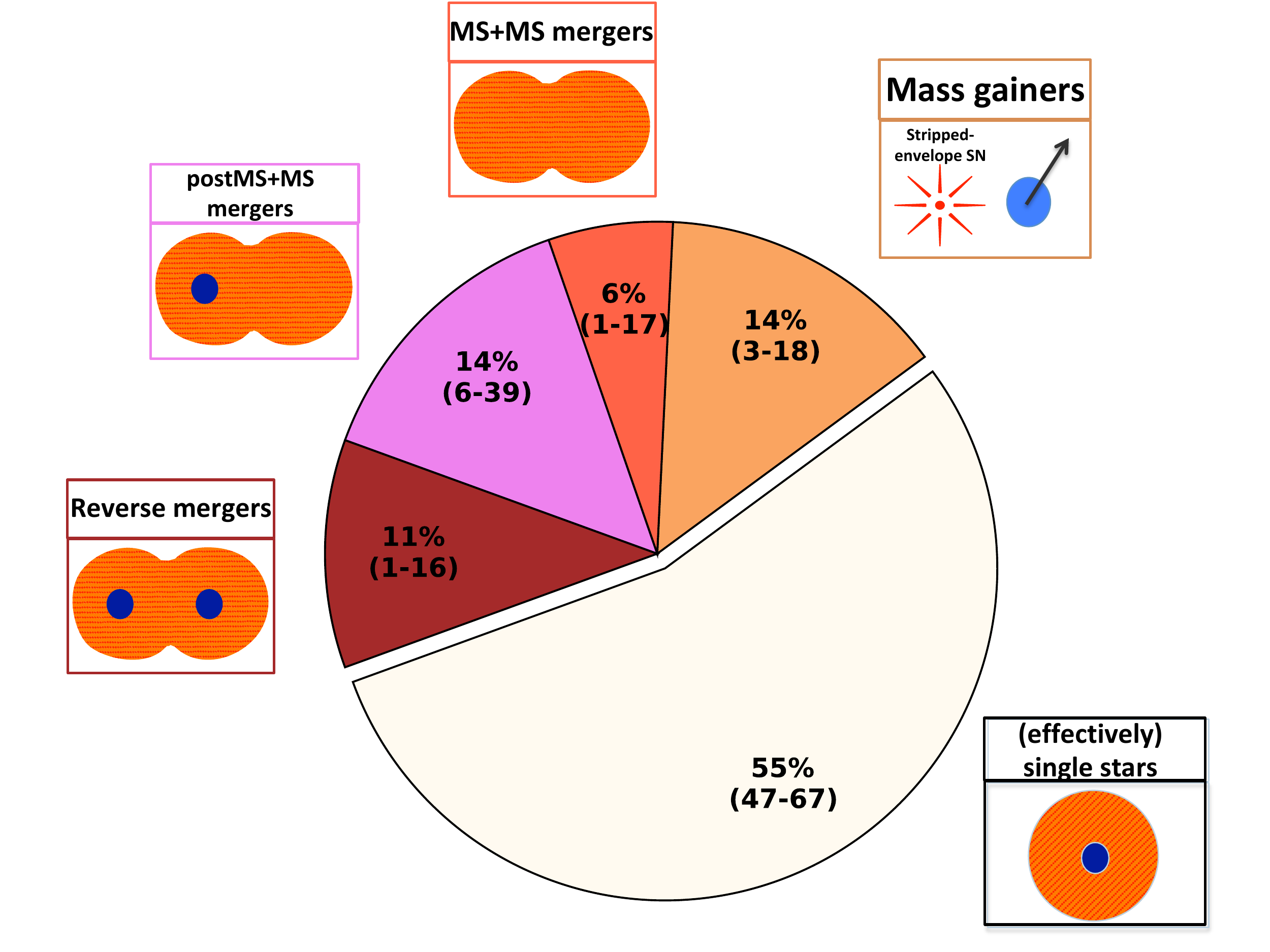}
\caption{Fraction of Type II SNe, assuming a realistic population in which 50\% of systems are binaries, that are expected to arise from stellar progenitors that have lived their lives as effectively single stars (white) and progenitors that have interacted with a binary companion through gaining mass or merging. See Section \ref{ch5:sec:overview_of_paths} for an overview of each evolutionary path.  
The main values correspond to the simulation with our standard assumptions whereas ranges in parenthesis refer to the upper and lower limits found in our parameter variations, excluding variations in binary fraction. 
}
\label{ch5:fig:piechart_typeII}
\end{figure*}

For binary stellar systems we account for the effect of tides \citep{Zahn1977,Hut1980,Hut1981}. During a RLOF mass transfer phase, we remove as much mass as needed from the donor star in order for it to remain inside its Roche lobe. We limit the accretion efficiency, $\beta$, not allowing the companion to accrete more than ten times its thermal rate, i.e. $M_{\mathrm{accretor}}/t_{\mathrm{Kelvin-Helmholtz}}$ \citep{Hurley+2002,Schneider+2015} in our fiducial model. We assume that the mass lost from the system carries specific angular momentum equal to that of the accreting star \citep[parameter $\gamma$;][]{van-den-Heuvel1994}. In our model variations we consider also extreme variations in $\beta$ and $\gamma$ to consider very conservative and highly non-conservative mass transfer, as described in \citetalias{Zapartas+2017}.
For the further evolution of mass gainers we assume that their internal structure adapts to their new mass as described in \citet{Braun+1995}, taking into account possible rejuvenation due to mixing of fresh material in the core \citep{Tout+1997,de-Mink+2013,Schneider+2015}.
%,Dray+2007

Unstable mass transfer or the swelling of the accreting star may lead to contact of the two stars and possibly CEE. 
We assume this to occur at the onset of RLOF for binary systems with a mass ratio more extreme than a critical value, $q_{\rm crit}$, which depends on the evolutionary phase of the star that is filling its Roche lobe. RLOF can occur when the donor is in its MS phase (where we assume  $q_{\rm crit,MS}\equiv0.65$), when the star is crossing the HG ($q_{\mathrm{crit,HG}} \equiv 0.4$, following \citet{Hurley+2002} and \citet{de-Mink+2014}), or when it is a giant-like star, where we follow the $q_{\rm crit}$ prescription of \citet{Hurley+2002}. This prescription is very simplified and we therefore consider large variations on these critical parameters. We assume that systems that come into contact during their MS evolution eventually merge, after losing a fraction $\mu_{\mathrm{loss}}\equiv0.1$ of the mass in the process and having $\mu_{\mathrm{mix}}\equiv0.1$ of hydrogen-rich material mixed in the core of the new star \citep{Lombardi+1995,Lombardi+1996,Gaburov+2008,de-Mink+2013}. If the donor has evolved off its MS, CEE is triggered. This process is treated according to the energy balance prescription of \citet{Webbink1984}, implementing the efficiency parameter for envelope ejection $\alpha_{\mathrm{CEE}}$, that represents the fraction of the lost orbital energy that is used in the ejection of the common envelope. We adopt the value of unity for this parameter in our standard assumptions, but consider variations up to one order of magnitude. The envelope mass distribution of the donor, which is needed to calculate its binding energy and is represented with the parameter $\lambda_{\rm CEE}$, is calculated following \citet{Dewi+2000,Dewi+2001} and \citet{Tauris+2001}. CEE can lead either to the ejection of the common envelope, leaving a binary system of tighter orbit, or alternatively to the merger of the two stars. The further evolution of a merger product is simulated as described in \citet{Hurley+2002} and \citet{de-Mink+2013}.

We compute the possibility of disruptions of binary systems due to mass loss during a SN \citep{Blaauw1961} and to the natal kick of the compact remnant. We assume random velocity natal kicks following a 1D Maxwellian distribution with a root-mean-square of $\sigma_{\mathrm{kick}} = 265 \, \mathrm{km}\,\mathrm{s}^{-1}$ \citep{Hobbs+2005}. In one of the model variations that we consider, we account for the possibility that some stars result in a direct collapse without producing a bright transient and thus are undetectable \citep[e.g,][]{OConnor+2011, Ugliano+2012}. Although the final outcome of a star is very sensitive to its initial mass \citep[e.g,][]{Sukhbold+2016}, in that variation we treat it in a more simplified way, assuming that all stars that form a core more massive than the one equivalent to a single star of $M_{\mathrm{max,ccSN}} = 20 \Msun $  initial mass do not produce a detectable event.

In order to account for the uncertainty in the minimum initial mass for ccSNe, $M_{\mathrm{min,ccSN}}$, we run simulations where we vary this parameter by changing the carbon-oxygen core mass threshold for a core-collapse accordingly, as in \citetalias{Zapartas+2017}.

For our standard simulation we assume that the initial mass $M_1$ of single stars as well as of the primary stars in a binary system, follow a \citet{Kroupa2001} initial mass function with a slope for massive stars of $\alpha_{\rm IMF}=-2.3$. The initial mass ratio distribution is flat, i.e. following a power-law of exponent $\kappa = 0$, between $0.1 < q = M_2/M_1 < 1$. The initial period distribution between $0.15 <  \log_{10} (P/\mathrm{day})< 3.5$ is assumed flat in the logarithmic space (with a power-law of exponent $\pi = 0$) for systems with initial primary mass $M_1<15$\Msun, whereas we follow \citet{Sana+2012} for higher mass binary systems. The latter favors short period systems ($\pi = -0.55$).
For our standard simulation we adopt a metallicity of $Z=0.014$. 

For our numerical simulations, we computed a grid of  $10^4$ single stars spread logarithmicaly in the initial mass parameter space, which ranges from $3$ to $100 \Msun$. For binary systems, in which the initial configuration is determined mainly by the initial masses of the two stars and the initial orbital period, we investigate the parameter space by computing $150\times 150 \times 150 \simeq 3.4 \times 10^6$ systems in the same parameter ranges as mentioned above, for our standard simulation.  For simulations where we vary our assumptions, the resolution of the grid is lowered by half for each of these three initial parameters.

The main difference between this work and \citetalias{Zapartas+2017} is that we make a more conservative assumption for the binary fraction of $f_{\mathrm{bin}}=0.5$ for most of our simulations. This is lower compared to the value 0.7 used in the grid of models of \citetalias{Zapartas+2017}, which however we still include as a variation. This is motivated by the empirical estimates of the binary fraction of early B-type stars that dominate the populations of  SNe\,II. %, more uncertain than in the case of  O-type stars, but 
These estimates seem to favor a lower value than in the case of O-type stars \citep[e.g.,][]{Duchene+2013,Moe+2017}.  We perform simulations for various binary fractions as well as for a simplified assumption of a mass dependent binary fraction, $f_{\rm bin}(M)$, as described in \citetalias{Zapartas+2017}.

In presenting our findings, the main values correspond to our fiducial simulation adopting a set of standard assumptions.  The uncertainty ranges shown refer to the largest differences found in our simulations when varying one-by-one the model parameters. % one at a time.
In the uncertainty ranges that we quote %in Figure~\ref{ch5:fig:piechart_typeII} and
in the rest of this study, we do not include the effect of different assumed binary fractions, although we consider them in our variations. The reason is that our findings by definition scale directly with the binary fraction. This usually dominates the uncertainty and thus obscures effects of the other parameter changes. We include the results of these simulations %in which we vary the binary fraction
in Models 45-47 of Table~\ref{ch5:table:parameters_uncertainties}.

\setlength{\tabcolsep}{2pt}
\begin{table*}[p] %p!
  \caption{Variations of the physical assumptions and initial conditions considered and their impact on  the fraction of each evolutionary channel for SN\,II progenitors and on the ratio of stripped-envelope to Type II SNe, $R_{\rm stripped/II} $. In the first column we link each varied parameter with the corresponding simulation of \citetalias{Zapartas+2017} whenever possible, although we assume $f_{\mathrm{bin}}=0.5$ as standard in this study. Results  from the simple analytical estimate in section \ref{ch5:sec:analytical_calculation} are also shown. \label{ch5:table:parameters_uncertainties}}
  \centering
   {\small
  \begin{tabular}{llrrrrrc@{\hspace{-10cm}}c@{\hspace{-10cm}}r}
  \hline \hline %inserts double horizontal lines
   \\[0.01ex]

{Model} &{Description} &  {effectively }& {mass}& {MS$+$MS  }& {postMS$+$MS  }& {reverse }& {spiraled-in } & {partially }  & {$R_{\rm stripped/II} $}\\
{in \citetalias{Zapartas+2017} }& &  {singles}        & {gainers}     & {mergers}  &   {mergers }         & {mergers} &  {during CEE} & {stripped} & \\
\hline
 %\\
 \\%[0.0015ex]
 &  & \multicolumn{7}{l}{[-------------------------------------------\%------------------------------------------]} &\\  
 %\\
 [0.05ex]
{\bf --} &  {\bf STANDARD ASSUMPTIONS$^{1}$     }                            &         {\bf 54.3} &     {\bf 13.7 }&    {\bf 5.6 }&    {\bf  13.9}  &   {\bf   11.3}  & {\bf 1.0} & {\bf  0.1} &{\bf 0.53} \\[0.5ex]
     \hline
\\     [0.5ex] 
{\bf --}&  {\bf simple analytical estimate (Section \ref{ch5:sec:analytical_calculation})	}  &         {\bf $\sim$ 56.0} &     {\bf $\sim$16.0} &       {\bf$\sim$ 8.5} &     {\bf $\sim$15.0 }&     {\bf $\sim$4.0} &     {\bf - }&      {\bf - } & {\bf $\sim$0.52}\\[0.5ex] 

\hline
\\
\multicolumn{6}{l}{\it \bf Physical assumptions} \\[0.5ex]
01 &  accretion efficiency $\beta=0$                                 &          57.6 &      13.8 &       5.8 &     14.6 &      7.1 &       1.1 &      0.1  & 0.49\\
02 &  accretion efficiency $\beta=0.2$                              &          55.3 &      17.6 &       5.4 &     14.1 &      6.6 &       1.0 &      0.1  & 0.49\\
03 &  accretion efficiency $\beta=1$                                 &          51.8 &      13.7 &       5.4 &     13.1 &     15.0&       0.9 &      0.0  & 0.61\\
04 &  angular momentum loss $\gamma=0$                       &          54.0 &      13.3 &       5.7 &     14.0 &     11.5&       1.4 &      0.1  & 0.53\\
05 &  angular momentum loss $\gamma$:circumbinary disk &          49.3 &        2.8 &       5.2 &     28.4 &     13.5&       0.8 &      0.0   & 0.34\\
06 &  merger $\mu_{\mathrm{loss}}=0$                            &         57.3 &      15.3 &       6.0 &       8.3 &     12.1 &       1.0 &      0.1  & 0.59\\
07 &  merger $\mu_{\mathrm{loss}}=0.25$                       &         54.1 &      13.7 &       5.7 &     13.9 &     11.5 &       1.0 &      0.1  & 0.52\\
08 &  merger $\mu_{\mathrm{mix}}=0$                             &         54.2 &      13.6 &       5.7 &     14.1 &     11.4 &       1.0 &      0.1  & 0.53\\
09 &  merger $\mu_{\mathrm{mix}}=1$                             &         54.2 &      13.7 &       5.6 &     13.9 &     11.6 &       1.0 &      0.1  & 0.53\\
10 &  no natal kick in SN, $\sigma_{\rm kick} = 0$               &         61.1 &        3.6 &       6.4 &     15.7 &     13.0 &       0.1 &      0.1  & 0.61\\
11 &  extremely high kick in SN, $\sigma_{\rm kick}=\infty$                &         53.7 &      14.0 &       5.7 &     14.1 &     11.4 &       1.0 &      0.1  & 0.53\\
12 &  $\alpha_{\mathrm{CEE}} = 0.1$                          		   &         47.4 &      12.0 &       5.0 &     19.4 &     15.9 &       0.3 &      0.1  & 0.42\\
13 &  $\alpha_{\mathrm{CEE}}= 0.2$                                  &         49.5 &      12.5 &       5.2 &     18.2 &     14.0 &       0.6 &      0.1  & 0.49\\
14 &  $\alpha_{\mathrm{CEE}} = 0.5$                                 &         55.4 &      14.8 &       5.8 &       9.0 &     14.1 &       0.9 &      0.1  & 0.56\\
15 &  $\alpha_{\mathrm{CEE}}= 2.0$                                  &         58.5 &      14.7 &       6.1 &     13.8 &      5.6 &       1.2 &      0.1  & 0.62\\
16 &  $\alpha_{\mathrm{CEE}} = 5.0$                                 &         63.7 &      16.2 &       6.7 &     10.1 &      1.6 &       1.6 &      0.1  & 0.69\\
17 &  $\alpha_{\mathrm{CEE}} = 10.0$                               &         67.1 &      17.1 &       7.0 &       5.5 &      1.1 &       2.0 &      0.1  & 0.75\\
18 &  $\lambda_{\mathrm{CEE}}=0.5$                                &         57.5 &      14.6 &       6.0 &     14.2 &      6.7 &       1.0 &      0.1  & 0.65\\
19 &  $q_{\mathrm{crit,MS}} = 0.25$                                &         54.3 &      13.7 &       5.4 &     13.9 &     11.7 &       1.0 &      0.1  & 0.53\\
20 &  $q_{\mathrm{crit,MS}} = 0.8$                                  &         54.1 &      13.6 &       7.0 &     14.1 &     10.3 &       1.0 &      0.1  & 0.53\\
21 &  $q_{\mathrm{crit,HG}} = 0.0$                     			   &         56.6 &      15.1 &       5.9 &       8.1 &     13.2 &       1.0 &      0.1  & 0.58\\
22 &  $q_{\mathrm{crit,HG}} = 0.25$                    			   &         57.3 &      15.2 &       6.0 &       8.3 &     12.1 &       1.0 &      0.1  & 0.58\\
23 &  $q_{\mathrm{crit,HG}} = 0.8$                     			   &         48.8 &        6.5&       5.1 &     30.8 &      7.7  &       0.9 &      0.1  & 0.41\\
24 &  $q_{\mathrm{crit,HG}} = 1.0$                     			   &         46.4 &        3.6 &       4.9 &     38.7 &      5.5 &       0.9 &      0.1  & 0.33\\
-    &  wind factor $\eta= 0.1$                                  				   &         56.2 &      15.5 &       6.9 &       8.2 &     11.6 &       1.1 &      0.5  & 0.37\\
25 &  wind factor $\eta= 0.33$                                  			   &         53.0 &      14.2 &       6.6 &     14.0 &     10.9 &       1.0 &      0.3  & 0.38\\
26 &  wind factor $\eta= 3.0$                                   				   &         56.0 &        9.3 &       4.0 &     14.6 &     15.2 &       0.9 &      0.0  & 1.35\\
29 &  exclusion of failed SNe ($M_{\mathrm{max,ccSN}}=20 \Msun$)&         54.6 &      12.7 &       4.7 &     14.5 &     12.4 &       1.1 &      0.0  & 0.30\\
%30 &$M_{\mathrm{min,ccSN}} = 1.30$ ($7.05$ init. mass) &         54.7 &      13.3 &       5.4 &     14.0 &     11.4 &       1.1 &      0.1 \\
%31 &$M_{\mathrm{min,ccSN}} = 1.40$ ($7.95$ init. mass) &         53.8 &      14.0 &       5.9 &     14.2 &     11.1 &       0.9 &      0.1 \\
30 &$M_{\mathrm{min,ccSN}} \sim 7 \Msun$ 						  &         54.7 &      13.3 &       5.4 &     14.0 &     11.4 &       1.1 &      0.1  & 0.48\\
31 &$M_{\mathrm{min,ccSN}} \sim 8 \Msun$ 						  &         53.8 &      14.0 &       5.9 &     14.2 &     11.1 &       0.9 &      0.1  & 0.58\\
%(see \citetalias{Zapartas+2017}) 

\\
\multicolumn{6}{l}{\it \bf Initial conditions} \\[0.5ex]

32 &  IMF slope for massive stars $\alpha_{\rm IMF}=-1.6$                 		  &         56.1 &     16.5 &       5.6 &     12.4 &      8.3 &      1.0 &      0.1  & 0.95\\ 
33 &  IMF slope for massive stars $\alpha_{\rm IMF}=-2.7$              	   	      &         52.9 &     12.2 &       5.7 &     14.7 &     13.5 &      0.9 &      0.1  & 0.41\\ 
34 &  IMF slope for massive stars $\alpha_{\rm IMF}=-3.0$             		      &         51.7 &     11.1 &       5.8 &     15.3 &     15.1 &      0.9 &      0.0  & 0.35\\ 
35 &  initial q distr. slope $\kappa=-1$                    			  &         54.6 &       8.0 &       6.7 &     24.1 &      5.9 &      0.5 &      0.1  & 0.42\\ 
36 &  initial q distr. slope $\kappa=+1$                    		  &         52.9 &     17.4 &       4.6 &       7.6 &     16.2 &      1.3 &      0.1 & 0.61 \\ 
37 &  initial period distr. slope $\pi=+1$                  			  &         58.1 &     17.5 &       0.8 &     12.8 &      8.6 &      2.0 &      0.2  & 0.60\\ 
38 &  initial period distr. slope $\pi=-1$                  			  &         47.0 &       8.6 &     16.8 &     11.3 &     15.9 &      0.3 &      0.0  & 0.41\\ 
39 &  $Z=0.0002$                                            				  &         56.4 &     13.6 &       1.2 &     14.1 &      8.0 &      1.6 &      5.0  & 0.32\\ 
40 &  $Z=0.001$                                             				  &         56.9 &     13.4 &       2.3 &     15.0 &      7.9 &      2.0 &      2.5  & 0.36\\ 
41 &  $Z=0.004$                                             				  &         56.8 &     13.7 &       4.1 &     14.3 &      8.9 &      1.7 &      0.5  & 0.39\\ 
42 &  $Z=0.008$                                             				  &         55.5 &     13.7 &       4.8 &     14.2 &     10.1 &      1.4 &      0.2  & 0.45\\ 
43 &  $Z=0.02$                                              					  &         56.7 &     12.9 &       6.7 &     13.9 &      8.5 &      1.4 &      0.0  & 0.67\\ 
44 &  $Z=0.03$                                              					  &         56.7 &     11.3 &       7.1 &     13.4 &     10.0 &      1.5 &      0.0  & 1.04\\

\\
\multicolumn{6}{l}{ {\bf Binary fractions}  \tiny{(not taken into account in the uncertainty range presented in this study)} }\\[0.5ex]
45 &  $f_{\mathrm{bin}} = 0.3$                            	         &         73.0 &       8.0 &       3.2 &      8.3 &      6.8 &       0.6 &       0.0  & 0.39\\ 
00 &  $f_{\mathrm{bin}} = 0.7$                            	         &         34.8 &     19.6&        8.0 &    19.9 &    16.2  &       1.5&        0.1 &0.67\\
46 &  $f_{\mathrm{bin}} = 0.99$                            	     &           5.3 &      28.2 &      11.8 &     29.1 &     23.6 &       2.0 &      0.1  & 0.88\\
47 &  mass dependent binary fraction,  $f_{\rm bin}(M)$                	  &         42.0 &      18.2 &       7.1 &     17.7 &     13.6 &      1.3 &       0.1  & 0.64\\ 
%[2.1ex]
   \hline \hline
  \end{tabular}
  } 
    \tablefoot{
\tablefoottext{1}{The standard assumptions of our numerical model are discussed in section \ref{ch5:sec:model}. In summary: $\beta$:$10*$(accretor's thermal rate),  $\gamma$:specific angular momentum of the accretor, $\mu_{\mathrm{loss}} = \mu_{\mathrm{mix}} = 0.1$,  $\sigma_{\rm kick}=265$ km $\rm{s}^{-1}$,   $\alpha_{\mathrm{CEE}} = 1.0$, $\lambda_{\mathrm{CEE}}$: \citet{Dewi+2000} \& \citet{Tauris+2001}, $q_{\mathrm{crit,HG}}=0.4$, $q_{\mathrm{crit,MS}}=0.65$ , $\eta = 1.0$, no failed SNe, $\alpha_{\rm IMF}=-2.3$, $\kappa=0$, $\pi=-0.55$ (for O-type primaries)  \& $0$ (for the  rest), $Z=0.014$, $f_{\mathrm{bin}} = 0.5$. }
}  
\end{table*}

\subsection{Numerical estimates of the SN relative rates and sensitivity to uncertainties}\label{ch5:sec:fractions}%\label{ch5:sec:fractions}

The results for our fiducial simulation following our standard set of assumptions as well as all the variations, are shown in Table~\ref{ch5:table:parameters_uncertainties}. For each simulation, we compute the expected fraction of SN\,II progenitors that follow any of the possible  evolutionary channels. 

Figure~\ref{ch5:fig:piechart_typeII} shows a graphical summary of the predicted fraction of SN\,II progenitors that have evolved as effectively single stars (white) or that have interacted with a binary companion (colored). The main fraction shown for each channel is the outcome for our set of standard assumptions whereas the range quoted in parenthesis depicts the minimum and maximum value found in our variations, assuming an initial binary fraction of  50\%.

We find that approximately half to two-thirds of SNe\,II, $47\%-67\%$ for all variations depicted in Figure~\ref{ch5:fig:piechart_typeII}, do not experience mass exchange by RLOF. This is either because they were born as single stars or because they have companions in very wide orbits, with  $P\gtrsim 1500$ days \citep[e.g.,][]{Yoon+2017}. 
Thus our computational results 
show that the slight majority of SNe\,II originate from progenitors  that lived their lives in isolation, even when binarity is taken into account. 

The remaining SN\,II progenitors (slightly below half for our standard assumptions) experienced mass exchange with their binary companions before explosion according to our simulations.  
This may include either mass accretion from the primary onto the companion star or merging of the two stars at some point during their evolution. Although the contribution of individual binary channels may significantly change in all model variations that we consider here (as discussed in the remainder of this subsection), the total fraction of progenitors that experience binary interaction is always significant in all model variations, i.e., never lower than about a 1/3 of all Type II SNe. 
The main effect of varying initial distributions and physical parameters is to affect the relative contribution of binary evolution channels producing SNe\,II with only a small effect on the total contribution. 

The fraction of SN\,II progenitors with binary interaction is only marginally dependent on most of the parameters considered in this work. The resulted fraction changes slightly on variation of parameters such as the slope of the IMF (Models 32-34), the mass accretion efficiency $\beta$ (Models 1-3), the strength of wind mass loss (parameter $\eta$, Models 25-25), and the metallicity $Z$ (Models 39-44).
The rate is sensitive to the fraction of close binary systems in our populations ($f_{\mathrm{bin}}$, Models 45 - 47 and  00 of \citetalias{Zapartas+2017} in Table~\ref{ch5:table:parameters_uncertainties}). A higher $f_{\mathrm{bin}}$ increases the fraction of SN\,II progenitors that experience mass gain or merge (and a corresponding decrease of the ones that lived as effectively single stars).

An increase in the total fraction of progenitors that experienced binary interaction and in particular those resulting from mergers after a common envelope phase (postMS+MS) is found in simulations where we have assumed a low value for the common envelope efficiency parameter $\alpha_{\mathrm{CEE}}$ 
(Models 12 and 13). A similar increase of binary progenitors occurs when we increase the number of systems that experience unstable RLOF, i.e., for high values of $q_{\mathrm{crit}}$. The increase is mainly due to the more likely occurrence of postMS+MS mergers, initiated during the HG phase of the donor star (Models 23 and 24). 
The opposite trend is found in simulation where we assume lower values of $\alpha_{\mathrm{CEE}}$, which results in more systems that successfully eject the envelope and avoid merging. This increases the number of stars that get stripped of their hydrogen-rich envelopes and thus do not produce SNe\,II.  An increase in the rate of SN\,II progenitors that experienced binary interaction is also observed for initial period distributions that favor binary systems with tight orbits (Model 38), as they result in mass exchange being even more probable.

In the remainder of this section we briefly discuss the main evolutionary scenarios %presented in Section \ref{ch5:sec:overview_of_paths}
and the sensitivity of their rates to variations of the assumed parameters. At the end of this section we also discuss two scenarios with minor contribution (listed as ``spiraled in during CEE" and ``partially stripped''), also shown in Table \ref{ch5:table:parameters_uncertainties}.

\paragraph{Mass gainers---}
Approximately 
$14 \% \, (3\%$ to $18 \%)$ of SN\,II progenitors 
accrete mass through RLOF from an initially more massive companion and are later ejected from the binary system when the companion explodes. %due to a prior explosion of their companion.
This prediction is particularly sensitive to assumptions concerning the evolution of the binary orbit during mass transfer and the SN kick of the primary.

This channel becomes less significant if we assume high angular momentum loss during the mass transfer phase from the primary, which results in more mergers or in tighter orbits after mass transfer (Model 5). Lower rates are also found in simulations in which a larger fraction of systems go through a CEE phase (high values of $q_{\mathrm{crit}}$, Models 23 and 24). We note that many of  these systems will still produce a Type II SN after possible merging. 
The rate of this channel is also low in the absence of a natal SN kick during the explosion of the primary (Model 10) because more systems stay bound. Such systems eventually experience a reverse mass transfer phase. In many cases we expect a stripped-envelope SN from the secondary star. 

A slight increase in the contribution of this channel to Type II binary SN progenitors occurs in our variations of initial distributions that favor massive secondaries, with masses similar to that of their primaries, and wider orbits (Models 36 and 37). This is mostly because the contribution of the merger channels decreases. 

The relative importance of this channel is not very sensitive to most of the remaining parameters we consider, such as the mass accretion efficiency $\beta$ (Models 1-3), the treatment of CEE ($\alpha_{\rm CEE}$ and $\lambda_{\rm CEE}$, Models 12-18), and the metallicity ($Z$, Models 39-44), and is only slightly sensitive to the slope of the IMF ($\alpha_{\rm IMF}$, Models 32-34).

\paragraph{MS+MS mergers---}
 About 
$31 \% \, (12\%$ to $44 \%)$ of all SN\,II progenitors in our simulations originate from merger products of two stars in a binary system. A subset of these,  
contributing about $6 \% \, (1\%$ to $17 \%)$ of all SNe\,II, come from a progenitor star that was the product of a merger between two main sequence stars.

The highest and lowest  occurrence values of this group correspond to initial distributions favoring shorter and wider orbits, respectively (Models 37 and 38). This is because MS+MS mergers can only occur in initially tight systems ($P\lesssim 5-10$ days). 
The rate of this channel is also lower in our models of lower metallicity, because in those cases stars are slightly more compact and expand less during their MS phase. It is hence less probable that they fill their Roche lobe and merge during this evolutionary phase. The rest of the parameters do not significantly influence the fraction of SNe\,II originating from this binary merger path.

We find that the contribution of this channel is not very sensitive to assumptions concerning the treatment of mass loss ($\mu_{\mathrm{loss}}$, Models 6 and 7 in Table~\ref{ch5:table:parameters_uncertainties}) and mixing ($\mu_{\mathrm{mix}}$, Models 8 and 9) for merger products.

\paragraph{postMS+MS mergers---}

Around $14 \% \, (6\%$ to $39 \%)$ of Type II SNe originate from %{\it mergers of a star that has evolved off the MS}
mergers  where the primary has evolved off the MS (postMS) before merging with its less massive, fairly unevolved MS companion. 

The fraction of Type II SNe arising from this channel is robust against variations of the mass accretion efficiency $\beta$ (Models 1-3), the slope of the IMF (Models 32-34), and the metallicity $Z$ (Models 39-44).
The prediction is sensitive to the number of binary systems that go through a CEE phase and the treatment of this process, i.e., the parameter $q_{\mathrm{crit,HG}}$, that controls how extreme the difference in mass of the two stars needs to be to trigger CEE. In case of the extreme assumption that all postMS mass transfer episodes lead to CEE (Model 24), the fraction of SNe\,II from these mergers reaches around $39\%$. This variation also leads to the highest total fraction of SN\,II progenitors that experience binary interaction, bringing it above half. Simulations of low $\alpha_{\rm CEE}$, that result in more mergers after the CEE phase, also increase the fraction of postMS+MS mergers. An increase is also observed for simulations of high angular momentum loss during the mass transfer phase (Model 5) because it leads to tight orbits and eventual merging.

Conversely, one of the lowest fractions of  postMS+MS mergers is found in our simulations in which we adopt an initial mass ratio distribution favoring companions with masses close to that of their primary stars ($\kappa = 1$; Model 36), because this increases the fraction of systems that undergo stable mass transfer.  Similar low rates for this channel are found for high values of $\alpha_{\rm CEE}$ (e.g., Model 10) which reduces the fraction of systems that merge during CEE.

\paragraph{Reverse mergers---}
In our suite of simulations, a significant fraction of $11 \% \, (1\%$ to $16 \%)$ of Type II SNe originate from binary systems where  reverse RLOF from the secondary towards the primary star triggers the merger.

The contribution of this scenario decreases for simulations in which the secondary stars have low relative initial masses (e.g., $\kappa = -1$; Model 35) or do not accrete enough mass from the primary during the first phases of mass transfer (i.e., for low accretion efficiency $\beta$; Models 1 and 2). This prevents the secondaries from having high mass at the moment of reverse mass transfer and reduces the possibility of scenarios that lead to SNe after the reverse mergers. In addition, this scenario becomes significantly less probable in cases that CEE results mostly in ejection of the envelope, avoiding the merger (e.g., $\alpha_{\rm CEE} >1$; Models 15 - 18).

\begin{figure*}[t]\center
\includegraphics[width=\textwidth]{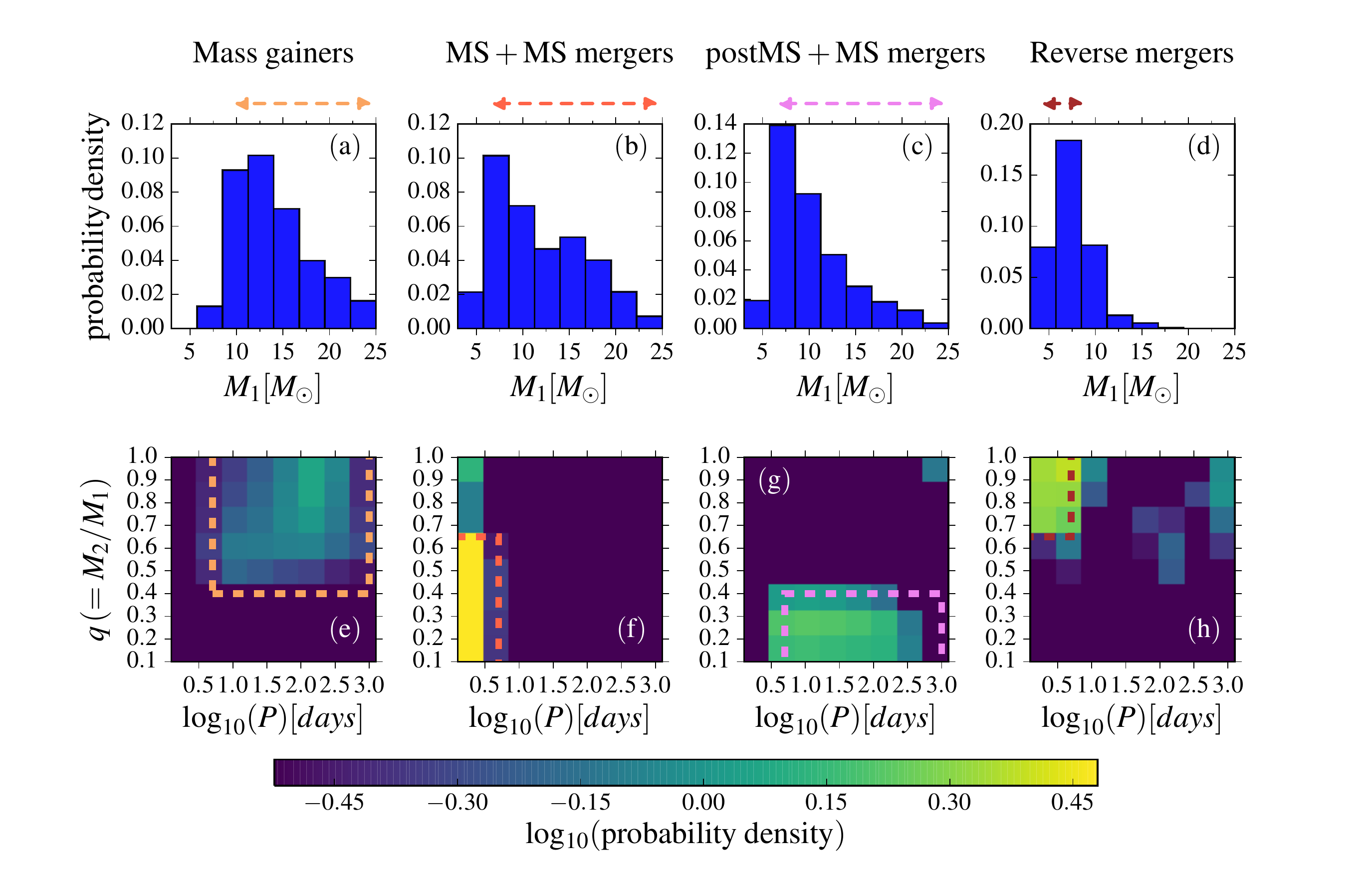}
\caption{The birth parameters of the progenitors that follow each main binary channel for SNe\,II. On the top row we show the probability density function of the initial primary mass, $M_{1}$, of the system and on the bottom the 2D density distribution of the mass ratio, $q=M_2/M_1$ and the orbital period, $P$, that these progenitors originate from. We also depict the range of the  initial primary mass (dashed arrows) and the assumed boundaries in the $P$-$q$ plane (dashed-line boxes, identical to Figure \ref{ch5:fig:parameter_space_cartoon}) that were assumed in the analytical estimate (section \ref{ch5:sec:analytical_calculation}).   
}
\label{ch5:fig:parameter_space_comparison}
\end{figure*}

\paragraph{Minority channels---}
A very small fraction, $1 \% \, (0\%$ to $2 \%)$, of SN\,II progenitors originate from the secondary stars in binary systems that experienced CEE during an unstable mass transfer episode from the initially more massive primary. They survive the {\it spiral-in} during the CEE phase, avoiding merging with the core of the primary. They are subsequently ejected from the system due to the SN of the primary and later produce a SN\,II themselves, similarly to the scenario of the mass gainers. The main difference is that, as CEE is a very fast phase  compared to the evolutionary timescale especially of the accretor \citep[e.g.,][]{Ivanova+2013}, we do not expect the progenitors to accrete a  significant amount of mass during spiral-in. Thus, the secondary stars that are  initially not massive enough for a SN, will still not be able to collapse even after CEE.  Another reason of the low contribution of this channel is that the orbit after the CEE is tight and the system cannot be easily disrupted during the first SN, thus the secondary cannot eventually produce a SN\,II.
The fraction increases when we assume high $\alpha_{\mathrm{CEE}}$  values or wide initial orbits (Models 17 and 37, respectively), in which cases the orbits after CEE are relatively wider and easier to get disrupted.

We also find a very small fraction of SN\,II progenitors from stars that initiated mass transfer towards a companion, either stable or unstable, but were only {\it partially stripped} of their hydrogen-rich envelope prior to explosion. The main reason for their low contribution is that, in our simulations, mass transfer usually leads to the almost complete removal of the hydrogen-rich envelope of the donor star, leaving a helium-rich core as a remnant and thus resulting in stripped-envelope SNe (possibly of Type IIb or Ib). 
We find that the fraction of partially stripped Type II SNe becomes non-negligible ($>1\%$) only for very low metallicities (e.g. Models 39 and 40).  
For predictions concerning this channel we believe it is better to refer to the results of studies that presented detailed calculation of the stellar structures \citep{Yoon+2017,Gotberg+2017,Eldridge+2018}.

\paragraph{Stripped-envelope SNe  ---}

We also compute the ratio of stripped-envelope SNe to all core-collapse events for all our simulations. In Table \ref{ch5:table:parameters_uncertainties}, we see that for our standard assumptions $ R_{\rm stripped / II} \sim 0.53$,  
which is very similar to our analytical estimate. In our population synthesis we can study the contributions of all the possible binary evolutionary channels to stripped-envelope SNe.  %not focusing only on the stipped primaries that experience stable mass transfer to the secondary. 
For our standard assumptions, we find that effectively single stars that get stripped through high wind mass loss only contribute to 1/4 of all stripped-envelope SNe. Primaries that experience stable mass transfer to the secondary have a similar contribution. The latter scenario is the only binary channel for stripped-envelope SNe considered in our analytical approach and is the same that also leads to Type II SNe from the  ``mass gainers'' secondary stars in case they are massive enough after their likely ejection. The rest of the hydrogen-poor events originate from other binary channels that in most cases  include either the ejection of the envelope of the primary during CEE, or a merger product that eventually gets stripped through its own wind mass loss, or the stripping of the secondary star during a reverse mass transfer that does not lead to merging. Thus, we see that although our analytical estimate had produced a similar value of $ R_{\rm stripped / II}$ to our numerical results, it could not  account for the wealth of binary channels that can lead to hydrogen-poor events.

The predicted value of $ R_{\rm stripped / II}$ is mainly sensitive to the assumptions for the stellar wind efficiency ($\eta$, e.g., Models 25, 26) and to metallicity $Z$ (Model 39-44). Higher wind mass loss rates lead to stripping of the hydrogen-rich layers of progenitors of lower initial mass, making the rate of stripped-SNe comparable to SNe\,II. Similarly, for high metallicity wind mass loss rates are enhanced, which has the same effect. Interestingly, even for inefficient winds and low metallicities the ratio does not drop below around 0.25 as binarity provides a mechanism for envelope stripping which is much less metallicity dependent and that dominates the stripped-envelope SN production in those conditions. A caveat here is that our simulations cannot treat partially stripping well, which can become important for low metallicities \citep{Yoon+2017,Gotberg+2017}. 

Low relative rates of stripped-envelope SNe are also found for assumptions that favor merging during CEE (e.g., high values of $q_{\rm crit,HG}$, Models 23, 24) or extreme angular momentum loss during stable RLOF and subsequent merging (Model 5). Simulations that disfavor high mass stars (Models 33, 34) or where we assume that no transient event is produced from these stars (Model 29) reduce the contribution of stripped-envelope SNe.

\subsection{Birth parameters of Type II SN progenitors and comparison with analytical estimate}\label{ch5:sec:comparison_with_analytics}

The fraction of expected SN\,II progenitors with prior binary interaction from our fiducial simulation, $\sim 45\%$, is in  good agreement with our analytical estimates in Section~\ref{ch5:sec:analytical_calculation}. 
The same holds roughly for the relative contribution of each channel, with the exception of reverse merger scenario, that we discuss below.  

In order to further understand the relative contribution of each channel we  study the initial binary configurations that lead to each scenario in our fiducial simulation, shown in Figure \ref{ch5:fig:parameter_space_comparison}. In the top panels the initial primary mass distribution is depicted. In all the channels apart from reverse mergers, the main contribution is coming from systems with initial primary mass of  around $\sim 10$ \Msun and declines for higher masses mainly due to the IMF (panels \emph{a-c}). We also see a significant contribution from masses $< M_{\rm min,ccSN} \approx 7.5 \Msun$ that is expected uniquely from binary systems due to mass exchange between the stars \citepalias[e.g.,][]{Zapartas+2017}. The latter is the dominant scenario for reverse mergers (panel \emph{d} in Figure \ref{ch5:fig:parameter_space_comparison}). The mass ranges that we assumed in our analytical estimate (colored arrows on top of panels \emph{a}-\emph{d}) are roughly consistent with our numerical results.

In the bottom panels of Figure \ref{ch5:fig:parameter_space_comparison}  we show the birth period and mass ratio parameters for each evolutionary scenario in our standard simulation (colored 2D histograms).  For comparison, we also show the boundaries followed in our analytical estimate (dashed-line boxes), the latter being identical to the boxes of Figure \ref{ch5:fig:parameter_space_cartoon}. The regions of origin of each scenario are in general agreement between our numerical and analytical approach. This agreement is partly expected because of some similar assumptions between our simulations and the analytical estimate, for example similar $q_{\rm crit}$. 
However, the birth regions in our numerical simulations have less clear boundaries in the parameter space and some differences with the simplified analytical boundaries are discussed below.

As expected, MS+MS mergers mainly originate from binary systems of short orbits and preferentially of unequal masses. According to our simulations, also stars in very tight binary systems of $\log_{10}P \lesssim 0.4$ ($\sim2.5$ days) cannot avoid merging regardless of the mass ratio, so even close to equal mass systems with $q > q_{\rm crit} = 0.65$ (top left corner of panel \emph{f} of Figure \ref{ch5:fig:parameter_space_comparison}).  
These systems experience very conservative mass transfer which initially shrinks the orbit. Thus, the secondary star, which gradually becomes more massive, eventually fills its Roche lobe too, leading to merging.

In our standard simulation we see that SNe\,II originating from mergers of an evolved star with a MS one can also result from initially very wide systems of almost equal mass components (top right corner of panel \emph{g} of Figure \ref{ch5:fig:parameter_space_comparison}). These are systems where the evolutionary timescale of the secondary star is similar to that of the donor, and thus it has also evolved off its MS at the start of CEE. 
The energy released during the spiral-in of the core of the secondary is not enough to unbind the much more massive common envelope, which consists of the two hydrogen-rich envelopes of the stellar components, and these systems tend to merge. This complicated evolutionary path has been ignored in our analytical calculations for simplicity and indeed the rate of this channel is low enough that it would not have changed the general results. In addition, we also find that the postMS+MS merger channel becomes less efficient for periods $\log_{10}P \gtrsim 2.0$ due to an increased probability of successful ejection of the common envelope in these systems, which we ignored for simplicity in our analytical estimate.

The discrepancy between the numerical and the analytical results in the rate  of the reverse merger channel is more significant ($\sim 11\%$ in our standard simulation compared to $\sim 4\%$ in our analytical estimate). Differences are also found in the birth parameters of this channel in the two methods. The assumption of our simple estimate that SNe\,II from reverse mergers mostly originate from close binary systems of stars with similar masses (that experience close to conservative mass transfer) is  consistent with our simulations, as seen in the top left corner of panel \emph{h} of Figure \ref{ch5:fig:parameter_space_comparison}. However, the regions in the parameter space in which these mergers lead to Type II SNe is more broad in our models, originating also from wider systems than assumed in the analytical approach. This is because reverse mergers are a diverse group involving the coalescence of two stars that can be in various stages of their evolution \citepalias{Zapartas+2017}. Reverse mergers from initially wide systems form massive enough cores to collapse, according to our models, although they have lost big part of their initial masses during mass transfer. They usually involve the merging of two stars almost at the end of their evolution, typically after the helium core exhaustion of the secondary.

\subsection{Comparison with previous theoretical studies}\label{ch5:sec:comparison_with_previous_work}

Previous studies have addressed the role of binarity on Type II SNe, either by presenting an overview of all the available binary channels for these events or by focusing on selected scenarios. 
In general we find that our results are in good agreement.

\citet[][hereafter \citetalias{Podsiadlowski+1992}]{Podsiadlowski+1992} were the first, as far as we know, to quantify the impact of different evolutionary channels that massive single and binary stars between $\approx 8-20$ \Msun may follow and to thereby estimate the implications for the diversity of core-collapse SNe.  
They estimate that around 37\% of systems that include at least one massive star will experience binary interaction (Figure 16 of \citetalias{Podsiadlowski+1992})\footnote{The fractions in Figure 16 of \citetalias{Podsiadlowski+1992} are normalized to the total number of stellar systems between 8-20 \Msun, not to the total number of Type II SNe. In principle, these two are not identical because some evolutionary channels may produce other Types of core-collapse supernovae (e.g., SNe Ib/c), avoid collapse, produce two Type II SNe from one system or lead to Type II SNe from other paths (for example from intermediate-mass binaries with $M_{1} \lesssim 8$\Msun). However, the values can be used for a rough comparison; this can be seen by the fact that the sum of the evolutionary paths that produce possible SNe\,II in their Figure 16 is still approximately 100\%.}. 
Our result of 
$\sim 46 \% \, (33\%$ to $53\%)$ of Type II SN progenitors that experienced binary interaction is similar to the value in \citetalias{Podsiadlowski+1992}. 
The main reason for this discrepancy is probably a different assumed fraction of stars which are in interacting binaries. \citetalias{Podsiadlowski+1992} assumes a binary fraction of $50\%$ and an initial period distribution that has a much wider range than assumed in our study. Such a distribution would be equivalent to a much lower $f_{\mathrm{bin}}\approx 0.3$ in our assumed orbital period range. 
A second reason is that we account for SNe originating from binary systems containing intermediate-mass primary stars with $M_1\lesssim M_{\rm min,ccSN} \approx 8$\Msun, which were not considered by \citetalias{Podsiadlowski+1992}.

Our predicted relative rates of individual evolutionary channels are also consistent with those of \citetalias{Podsiadlowski+1992}.  Specifically for the SNe\,II that come from MS+MS mergers (a few \% in both studies), from secondary stars that survived a common envelope evolution (around 1\%), and from primary stars that were partially stripped (2\% in their work, $\lesssim$ 1\% in this study). We predict slightly more SN\,II progenitors that gain mass from a companion ($\sim 14\%$ compared to $\lesssim 10\%$ in their work), as \citetalias{Podsiadlowski+1992} do not account for supernova kicks 
 that can unbind the system. Without system disruption the secondary star could lose its hydrogen envelope during a phase of reverse mass transfer, and so not produce a Type II SN. 
We also predict a higher fraction of SNe\,II from mergers of a postMS star with a MS companion ($\sim$ 14\% compared to $\sim$ 4\% in their case). This is mostly due to different assumptions in the initial mass-ratio distribution and in the treatment of CEE. Finally, our assumptions lead to predicting a significant fraction of Type II SNe from reverse mergers; the majority of these originate from systems with initially intermediate-mass stars. \citetalias{Podsiadlowski+1992} do not consider this evolutionary path. Part of their predicted 14\% of systems that lead to a WD/SNIa from the primary may produce a SN\,II if it merges with its companion.

 \citet{Vanbeveren+2013} focus on the expected rates of mergers during the HG phase of the donor (specifically, when the donor star still has a radiative envelope). Some of these may produce BSG SN\,II progenitors, as might some from our category of postMS+MS mergers. For similar assumptions (e.g., $\alpha_{\rm IMF}=2.3$, $\alpha_{\rm CEE} = 1$) they find that roughly one third to half of binary systems will initiate mass transfer during this phase. A large portion of them will eventually merge, although they show, as we do, that this outcome is sensitive to model assumptions. If we adopt their computed rate for these mergers to crudely estimate the predicted fraction of SNe\,II that are postMS+MS mergers, for $f_{\rm bin}=0.5$, we find their values, close to 20-25\%, are roughly consistent with our predicted $14\% (6-39\%)$ fraction.

The fraction of SN\,II progenitors our calculations predict to have gained mass and then been ejected from their binary system as runaway stars is very similar to that found in \citet{Eldridge+2011}. Their Table 11 states that slightly more than 10\% of SN\,II progenitors are predicted to follow this scenario, in good agreement with the rate from our standard model ($15\%$ of SNe\,II). 
 
\citet{Eldridge+2018} have modelled the light curves of Type II SNe from binary-star models.  They also conclude that binarity plays a crucial role in the diversity of Type II SNe.   Binary interactions naturally influence the variety of light curves, in large part by altering the final masses of the hydrogen-rich envelopes of the SN progenitors.  \citet{Eldridge+2018} also estimate the fraction of each light-curve type (e.g. Type II-P, II-L, etc) from their model population. However, as they mention, their population synthesis calculation is sparse and they only take into account the primary star of each system.  Our simulations predict a significant contribution to the Type II SN population from mass-gaining secondary stars or reverse-merger progenitors.

We do not simulate the light curves from the binary products \cite[as done in][]{Eldridge+2018}. 
However, if we make broad assumptions about the dominant light-curve type from progenitors in each evolutionary channel, as discussed later in Section \ref{ch5:sec:diversity_type} and summarized in Table \ref{ch5:table:channel_properties}, their fractions of SN types are plausibly consistent with our results.  One of the differences is that many binary systems in their study lead to partially or fully stripped progenitors (mostly ending up as Type IIb SNe) from channels that we predict to result in postMS+MS mergers and thus to produce SNe with more massive hydrogen-rich envelopes; this is mainly due to a different treatment of mass-transfer stability in the two codes.

\begin{figure}[t]\center
\includegraphics[width=0.5\textwidth]{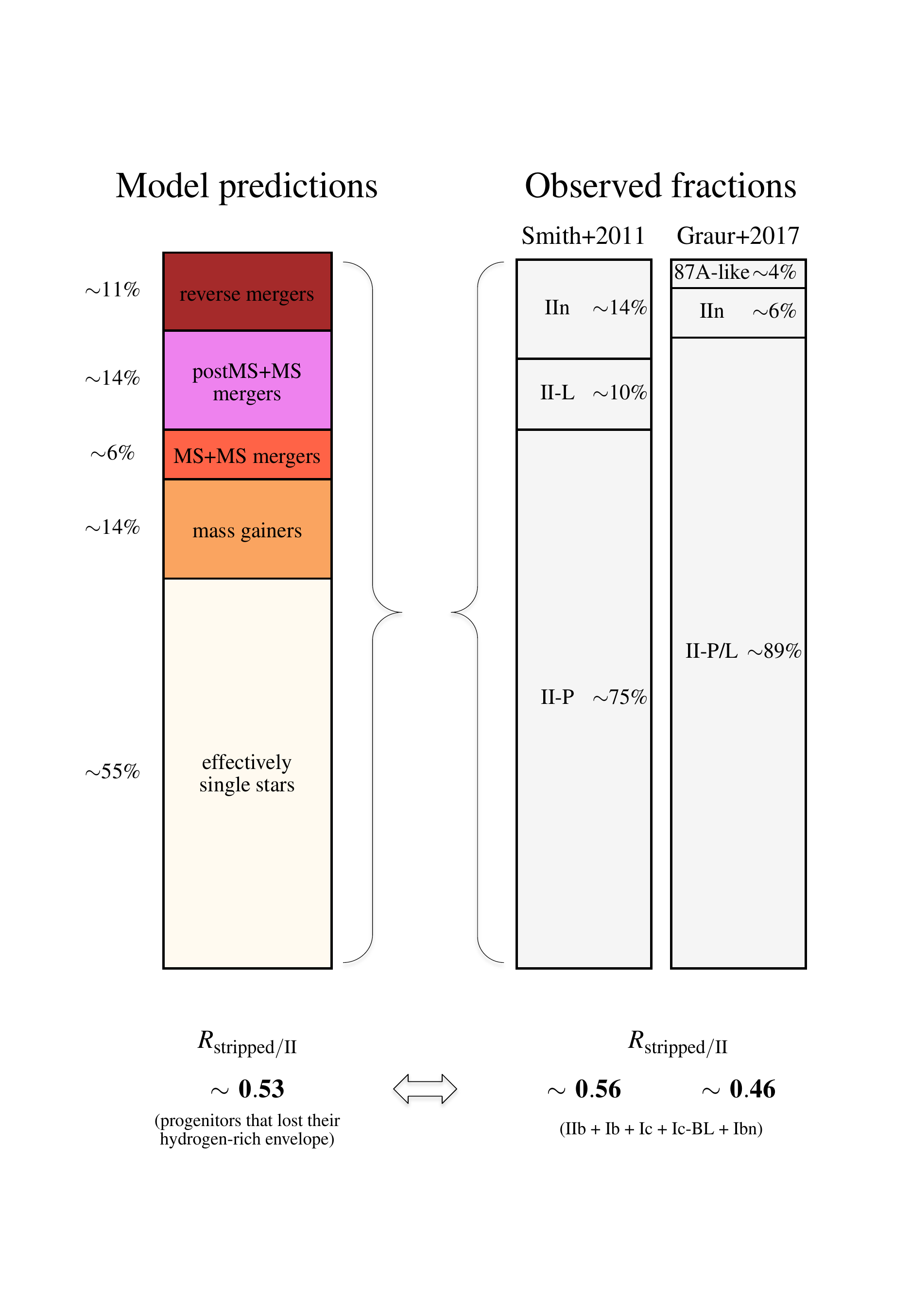}
\caption{Comparison of theoretical predictions in this work for the fraction of each SN progenitor scenario (left column) with observationally-inferred rates (from \citealt{Smith+2011} and  \citealt{Graur+2017a}, as labelled) of each hydrogen-rich SN type. The ratio of stripped-envelope to Type II SNe is also shown at the bottom.
}
\label{ch5:fig:observational_piechart}
\end{figure}

\section{Discussion}\label{ch5:sec:comparison_with_observations}

\subsection{Global comparison with observational rates of SNe~II}\label{ch5:sec:global_comparison}

Binarity adds a variety of evolutionary scenarios towards Type II SN progenitors. 
We do not expect that there is a one-to-one mapping between all progenitors from a broad binary evolution channel and an observed hydrogen-rich SN  subtype.  However, we consider it worth placing our predictions about the relative rates of each evolutionary path in the context of the observed fractions of the subtypes of Type II SNe.

Several groups have investigated the observational rates of Type II SNe from transient surveys \citep[e.g.,][\citealt{Graur+2017a}, which is a revisit of the data presented in \citealt{Li+2011}]{Smartt+2009,Smith+2011}.  Here we compare to the bias-corrected sample from Figure 1 of \citet{Smith+2011}, and the overall volume-limited sample from \citealt{Graur+2017a}.   From these, we select only the hydrogen-rich SNe as defined in our study: Type II-P, II-L and IIn SNe (i.e., not including Type IIb).

Figure~\ref{ch5:fig:observational_piechart} compares the relative observed rates for SNe\,II with the predicted fractions of each evolutionary channel for Type II SNe (left column).  The general comparison indicates that binary interactions should significantly contribute to determining the diversity seen in Type II SN populations.

Even the population of ``normal'' Type II-P SNe contains byproducts of binary evolution.  Type II-P SNe, the most abundant SN subtype (see Figure~\ref{ch5:fig:observational_piechart}), and their RSG progenitors \citep[see, e.g.,][]{Smartt+2009} are usually assumed to have experienced a ``normal'' single-star evolution (e.g., \citealt{Arcavi2017} and references therein).  Comparing the observed rates to our theoretical predictions indicates that the fraction of SN\,II progenitors that lived effectively as single stars is insufficient to account for all observed Type II-P SNe. Part of the Type II-P SNe progenitor population has experienced some kind of binary mass exchange before exploding. For our standard assumptions this would be roughly 1/3 of all Type II-P SNe. The exact value is subject to the model uncertainties (discussed in Section \ref{ch5:sec:fractions} and Table \ref{ch5:table:parameters_uncertainties}), as well as to uncertainties in the empirical rates.  However, according to our models, Type II-P SNe could only originate entirely from non-interacting stars in stellar populations with an unrealistically low fraction of interacting binary stars, around $f_{\mathrm{bin}} \sim 0.3$.

 \begin{table*}[t]%[p] %p!
  \caption{Summary of speculative properties for the Type II SN progenitors originating from each evolutionary channel.  \label{ch5:table:channel_properties}}
  \centering
   {\small
  %\begin{tabular}{lllllllll}
  \begin{tabular}{l||c|c|c|c|c}
  %\hline \hline %inserts double horizontal lines
% &\multicolumn{5}{l}\\
   %\\[0.01ex]
             &  (effectively) & mass & MS$+$MS  & postMS$+$MS  & reverse  \\
 & single stars&  gainers           & mergers  &   mergers          & mergers\\  %Property
\hline \hline
Computed & 55\%  & 14\%& 6\%& 14\%& 11\%\\
fraction $^{(a)}$ &  (47\%-67\%)  & (3\%-18\%) & (1\%-17\%) & (6\%-39\%) & (1\%-16\%) \\
 \hline
% \\[0.1ex]
SN Type $^{(b)}$ & II-P, II-L?  & II-P, IIn?  & II-P, IIn?  & II-P?, 87A-like?, IIn?, & II-P?, IIn??, II-L?\\
 &   &  &  &  II-L??, superluminous?? & \\
  \hline
CSM of binary origin  & no  & no & no  & no?  & maybe \\
at explosion $^{(c)}$ &  & &  &  &  \\
 \hline
 Surrounding population & normal & far away & older & older & possibly\\
 of progenitor $^{(d)}$ &  			&  (ejected)			&       	&					&much older\\
 \hline
 Binary companion & no  & no  &no  & no  &  no\\
at explosion $^{(e)}$& (unless in wide binary)  & (ejected)  & (merged)  &(merged)  &(merged) \\
 \hline
Rotation, & birth& enhanced & enhanced rotation, & ? &  ?\\
 magnetic field  $^{(f)}$ &  &rotation & (\& magnetic field?)   &  &  \\
 \hline
 
  \end{tabular}
  }
  \tablefoot{
  Discussed in Section: 
\tablefoottext{a}{\ref{ch5:sec:fractions} (Figure \ref{ch5:fig:piechart_typeII})}, 
\tablefoottext{b}{\ref{ch5:sec:diversity_type}}, 
\tablefoottext{c}{\ref{ch5:sec:CSM}},
\tablefoottext{d}{\ref{ch5:sec:environment}}, 
\tablefoottext{e}{\ref{ch5:sec:no_companions}}, 
\tablefoottext{f}{\ref{ch5:sec:overview_of_paths}.}
}  
\end{table*} 

\subsection{Speculations on subtypes from binary evolutionary scenarios and comparison with rates}\label{ch5:sec:diversity_type}

Similar binary evolution paths could result in qualitatively different Type II SN subtypes depending, e.g., on the mass ratio or timing of a merger, or on different amounts of mass lost or accreted in mass-transfer phases. With this caution in mind we now speculatively discuss potential SN diversity from each evolutionary scenario.   The summary of this discussion is shown in Table \ref{ch5:table:channel_properties}, among other potential properties of each scenario.   We note that it is debated whether Type II-L SNe should be treated as a clearly distinct subtype from Type II-P SNe (compare, e.g., the two observational samples in Figure~\ref{ch5:fig:observational_piechart}).

 \subsubsection{Envelope masses at explosion} \label{ch5:sec:envelope}

Some of the observed diversity in Type II SNe is a consequence of the mass of the hydrogen-rich envelope at explosion.  Binary interactions can drive the envelope mass fraction at explosion to be different than what is expected in single-star evolution.  Lower hydrogen-rich envelope masses make it harder for the characteristic plateau of Type II-P SNe to form (see, e.g., \citealt{Popov1993}). When assuming this explains the difference in light-curve shapes, a plausible hydrogen-rich envelope mass dividing Type II-L from Type II-P SNe is $\approx 2 \Msun$ (see, e.g., \citealt{Heger+2003}).   For sufficiently high envelope mass fractions stars are blue supergiants at the time of explosion (see Section \ref{ch5:sec:overview_of_paths}).  

Single stars, or effectively single stars, which explode as hydrogen-rich SNe seem overwhelmingly likely to explode with sufficiently massive hydrogen-rich envelopes to produce Type II-P SNe.  However we cannot exclude that, e.g., mass loss through winds could mean that some explode without producing a plateau in their light curve.   We do not expect effectively single stars to explode as BSGs, unless we assume  extreme physical parameters \citep[e.g.,][]{Langer1991a}.

Stars that accrete mass during their MS, or merger products of MS+MS stars, rearrange their structures to resemble those of more massive single MS stars (see Section \ref{ch5:sec:overview_of_paths}).  Hence, to first order, it seems reasonable to expect that MS mass gainers and mergers produce a similar diversity of hydrogen-rich SNe as (effectively) single stars.  However, it may well be that the post-interaction spin angular momentum sometimes leads to a significant effect on the SN.  Also, post-merger and post-accretion structures are not always identical to true single-star structures, e.g., regarding the helium profile outside the convective core.

Comparing to Fig. \ref{ch5:fig:observational_piechart}, the evolutionary channels described in the previous two paragraphs would be roughly enough by rate to explain the fraction of Type II-P SNe given by \citet{Smith+2011}, but not the combined rate of Type II-P and Type II-L SNe.

Some postMS+MS mergers and reverse mergers should also contribute to the rate of Type II SNe with a relatively normal range of envelope masses. Thus, they may produce similar observational signatures with single star progenitors.  However, merger products can lead to peculiar envelope mass fractions, depending on how much mass they lose during their CEE phase. A subset of postMS+MS mergers are expected to increase the final envelope mass fractions enough to produce a BSG SN progenitor at explosion, and thereby to a SN 1987A-like explosion \citep[e.g.,][]{Podsiadlowski+1990,Podsiadlowski1992,Menon+2017}. If no reverse mergers contribute to the rate of SNe from exploding BSGs, then Fig. \ref{ch5:fig:observational_piechart} suggests that roughly less than a third of postMS+MS mergers produce BSGs which explode as a SN 1987A-like event.

We cannot exclude that postMS+MS and reverse mergers sometimes lead to net loss of material from the hydrogen-rich envelope before re-adjustment of the merged star (e.g., after a CEE phase with most of the hydrogen-rich common envelope ejected).   This possibility might help to explain how those channels could contribute to the Type II-L SN rate.   

One channel occurs too rarely in our models to be shown in Figure~\ref{ch5:fig:overview_cartoon} and in Table~\ref{ch5:table:channel_properties}.  A small fraction of our SN progenitors are predicted to retain a hydrogen-rich layer after stable RLOF onto a companion.  These partially stripped progenitors may lead to SNe with features resembling Type II-L or Type IIb SNe \citep[e.g.,][]{Yoon+2010,Claeys+2011,Morozova+2016}.

\subsubsection{Circumstellar material} \label{ch5:sec:CSM}

Type IIn SNe display clear evidence for interaction with nearby hydrogen-rich circumstellar material (CSM).  It has also been argued that the physical conditions that lead to Type II-L SN lightcurves may not normally be due to progenitors with relatively low-mass envelopes but instead result from the occurrence of CSM that is less dense than in the case of Type IIn SNe \citep[e.g.,][]{Morozova+2017}. In that case, Type II-L SNe could be seen as less extreme cases of interaction with CSM as compared to Type IIn SNe.    In addition, some Type II-P events show transient observational features of CSM interaction soon after explosion \citep{Smith+2015a,Khazov+2016,Yaron+2017}.

If this CSM is a consequence of binary interactions, then in principle the CSM might be leftover matter from a late mass-transfer episode or merger \citep[see, e.g.,][]{Smith+2014}.   Our population of reverse mergers are predicted to have the shortest typical time difference between the merger and the SN compared to all other channels, of the order of $10^5$ years for roughly half of the reverse-merger progenitors in our simulations.  However, that typical time difference is still very long compared to the empirically-inferred time-lag between the mass ejection and the explosion for known Type IIn SNe, which is of the order of 10 to 100 years \citep{Smith2017}. On the other hand, binary interaction may potentially be the origin of remaining material in the vicinity of the progenitor, but not as close in distance as in type IIn SNe. This has been suggested for example as an explanation of the ring structure around the SN1987A progenitor \citep{Morris+2007}. 

The likelihood of binary interaction during just the final few years or decades before the SN event is very low, unless the interaction is caused by the immediate pre-SN evolution of the star.  Inflation of the stellar envelope may indeed occur during the last phases of evolution, perhaps because of nuclear burning instabilities \citep[e.g.][]{Mcley+2014,Smith+2014}, or because of energy deposited from waves excited near the core \citep{Fuller2017}.    Alternatively, these internal stellar processes have been suggested as the main intrinsic cause of the CSM production, independently of binary interactions. Such CSM production due to internal processes or fine-tuned late expansion of a star close in time with the SN event is not included in our simulations.

Binary interactions might also lead to a stellar product which later ejects sufficient CSM to lead to observable consequences during the SN event, for example SN progenitors which are Luminous Blue Variables \citep[LBVs,][]{Smith+2007,Smith+2015}.   \citet{Justham+2014} find that suitable mergers and mass gainers could be LBV-like stars immediately before core collapse, possibly leading to a subset of Type IIn SNe.  However, the specific mergers discussed in \citet{Justham+2014} are not in our predicted population, due to different assumptions about mass-transfer stability.

\subsubsection{Thermonuclear SNe from some reverse mergers}

Typical observational signatures of explosions arising from the reverse merger channel are difficult to state confidently.  The evolution of such merger products is highly uncertain.   Some reverse mergers in which the primary formed a white dwarf prior to coalescence might perhaps produce thermonuclear explosions, similar to a Type Ia SNe, rather than core-collapse SNe.   However, in contrast to a canonical Type Ia SN, the progenitor would be surrounded by hydrogen-rich layers  \citepalias[for further discussion of this channel see][]{Zapartas+2017}. Some observed SNe which show Type IIn features are already believed to have been powered by thermonuclear explosions, such as SN 2002ic and SN 2005gj, usually called Type Ia-CSM \citep{Hamuy+2003,Kotak+2004,Silverman+2013}, although other explanations arise naturally from some existing Type Ia SN models \citep[e.g.,][]{Han+PhP2006,WangJustham+2014}.

\subsubsection{Metallicity dependence}

The different predicted evolutionary channels are not particularly sensitive to the metallicity, $Z$ (models 39-44).   The overall fractions of only two predicted evolutionary channels are sensitive to metallicity.  First, we find that MS+MS mergers are less likely at lower metallicity, because the stars are less prone to interact during their main-sequence phase.  Second, the production of SN progenitors which are partially stripped by stable mass transfer is more likely at low $Z$, because the progenitors are more likely to retain their remaining post-mass-transfer hydrogen-rich envelopes, due to lower rates of wind mass loss.   Our models may well underestimate the true metallicity dependence of this partial stripping \citep[see, e.g.,][]{Yoon+2017,Gotberg+2017}.

Empirically, no significant dependence of Type II SN subtypes on the metallicity of the host galaxies is found in the Lick Observatory Supernova Search \citep{Graur+2017, Graur+2017a}.  However, 1987A-like SNe have so far been found exclusively in low-metallicity host galaxies \citep{Taddia+2016,Graur+2017a}.  If this is not a statistical accident from the small sample of discoveries, this might potentially point to 87A-like SNe originating from a specific class of postMS+MS mergers that is metallicity dependent, or that the progenitors need low metallicity conditions to retain their structure as BSGs until their collapse.

\subsection{Rate of stripped-envelope SNe compared with observations}
In Figure \ref{ch5:fig:observational_piechart} we also compare the ratio of stripped-envelope SNe to Type II events. 
In both our analytical and our standard numerical results we find a value for the relative rate of stripped-envelope SNe compared to Type II events of roughly $R_{\rm stripped/II} \sim 0.53$.

Our simulations find that this ratio is quite robust to model assumptions, although it is dependent on metallicity, wind mass-loss efficiency, the possibility of no transient from very massive progenitors and the slope of the IMF (see Section \ref{ch5:sec:fractions} and Table \ref{ch5:table:parameters_uncertainties}). This value is consistent with the value of 0.56 and 0.46 inferred from observations by \citet{Smith+2011} and \citet{Graur+2017a}, respectively, shown in Figure~\ref{ch5:fig:observational_piechart}.  It is also roughly consistent with the observed relative rates from other studies too \citep[e.g.,][find the ratio of $R_{\rm stripped/II}$ to be around 0.61 and 0.43, respectively, where we have included the events classified as Type IIb, Ib and Ic in these studies as stripped-envelope SNe]{Eldridge+2013, Shivvers+2017}.

\subsection{Surrounding environment and corresponding age of Type II SNe} \label{ch5:sec:environment}

SN\,II progenitors from a population that includes binary systems are on average expected to be part of older stellar populations than from the equivalent single star progenitors. The main reason for this is because progenitors that  experienced mass transfer or merging have had typically a longer lifetime relative to the single star progenitors of similar final mass, leading to delay-times than cannot be reproduced by massive single stars \citep[$\gtrsim 50$  Myrs up to 200 Myrs in extreme cases;][\citetalias{Zapartas+2017}]{De-Donder+2003a}. 

The longer delay-times also imply that the progenitors have more time to drift away from their birth location. This should be especially important for the channel of reverse mergers, in which the evolutionary lifetimes of the usually intermediate-mass components of the binary system are the longest prior to merging, of the order of tens of Myrs more than single star progenitors  \citepalias{Zapartas+2017}. The fact that one of the possible outcomes of this path may display signatures of CSM interaction (see Section \ref{ch5:sec:diversity_type}) seems consistent with observational evidence that finds Type IIn SNe in low correlation with star-forming regions, pointing to lower mass progenitors than what was initially believed \citep{Anderson+2012,Habergham+2014,Kuncarayakti+2018}. 

On top of that, Type II SNe from mass gainers may have been ejected from their birth place due to disruption of the binary system. These progenitors can thus be  associated with a surrounding population that has characteristics that are not typical for star-forming regions (e.g., \citealt{Renzo+2019} find an average travel distance of ejected mass gainers until explosion of $\gtrsim 100$ pc). 
The scenario of binary ejection has been proposed to explain the presumed isolation of LBV stars \citep[][although see also  \citealt{Humphreys+2016} and \citealt{Smith2016}]{Smith+2015}, which are potential progenitors of Type IIn SNe \citep{Gal-Yam+2007,Smith+2015}. 

Observations of the surrounding population of Type II SNe or of SN remnants can provide constraints on the evolutionary channels of their progenitors \citep[e.g.,][]{Murphy+2011,Williams+2014,Jennings+2014,Maund2017,Diaz-Rodriguez+2018,Xiao+2018,Auchettl+2018}. 
The tighter correlation of stripped-envelope SNe, such as Type Ib and Ic, with star-forming regions compared to SNe\,II \citep{Anderson+2012,Habergham+2014,Kuncarayakti+2018} can be interpreted as an indication that  the former originate from a distinct mass range of higher-mass progenitors, possibly losing their envelope through winds.  
However, the fact that SNe\,II from binary systems are naturally expected to be found in older populations and possibly far away from their birth places, leaves room for an alternative interpretation in the context of stripped-envelope SNe originating predominantly from interacting binary progenitors. 
This is because, even if most stripped-envelope SN progenitors lose their envelope during RLOF and thus originating from a wider initial mass range than single Wolf-Rayet stars, they are still expected to have {\it on average} relative shorter delay-times and tighter correlation with star-forming regions compared to SNe\,II \citep[][\citetalias{Zapartas+2017}]{De-Donder+2003a,Xiao+2015}.

\subsection{No stellar companions expected for Type II SNe}\label{ch5:sec:no_companions}
Our findings imply that the vast majority, if not almost all SN\,II progenitors, should not have a companion star in their vicinity at the moment of explosion. This is a consequence of the fact that all the dominant binary scenarios for SNe\,II involve either the merging of the two stars or, in the case of mass accretors, the ejection of the progenitor from the binary system due to a prior explosion. 
According to our simulations, only ``effectively single'' SN\,II progenitors that are in very wide binary systems \citep[typically of $\gtrsim 1500$ days period; e.g.][]{Sana+2012} will have a binary companion at the moment of explosion. 
This would imply that in case a companion is detected at the SN site of a Type II event, it has never exchanged mass with the progenitor star.

The exceptions are SN\,II progenitors that are partially stripped due to mass transfer onto a companion that is still there at the time of explosion, but this channel is found to be negligible for our standard assumptions.
We also do not consider here the quite significant probability of the progenitor being a member of an initial triple or even higher-order system \citep{Eggleton+2008,Duchene+2013,Sana+2014,Moe+2017}. In these cases, without considering the possible complications on the evolution of the inner binary system due to the third star \citep{Toonen+2016}, merger channels that are discussed in this study  would still have that star nearby at the time of explosion. For the cases of mass gainers, the fact that the progenitors will be ejected due to a prior explosion excludes the possibility of a nearby companion anyway. 

The prediction of a fairly high absence of companions at the moment of a SN\,II event is considerably different compared to the case of stripped-envelope, hydrogen-poor SNe (Type IIb, Ib, Ic, Ic-BL, Ibn). \citet{Zapartas+2017a} made theoretical predictions about the companions of hydrogen-poor explosions, using the same code and very similar input assumptions. They find that the majority of stripped-envelope SNe are expected to have a companion at the moment of explosion, predominantly an unevolved star during its MS phase, which usually was the cause of the stripping of the SN progenitor.  

There have been observational searches for companions of hydrogen-rich SNe. 
So far, none of these searches for companions of Type II SNe resulted in a detection, establishing upper limits on possible companions for the Type II-P SN 2005cs \citep{Maund+2005, Li+2006}, the Type II-P SN 2008bk \citep{Mattila+2008} and for the Type II-peculiar SN1987A \citep{Kochanek2017}. In addition, the Crab, which was probably a Type II SN, seems to lack a surviving companion \citep{Kochanek2018}. Thus, our predictions are consistent with the so-far lack of companion detections in cases of SNe\,II and may pinpoint the reason for it.

\section{Summary and Conclusions}\label{ch5:sec:summary}

The aim of this study is to investigate the diversity of single and binary stellar evolutionary channels that Type II supernovae (``SNe\,II'', including Type II-P, Type II-L, Type IIn and possibly Type II-peculiar) progenitors may follow.

\begin{itemize}
\item A significant fraction of SN\,II progenitors, roughly between $1/2$ and $1/3$, are expected to interact with a companion before exploding. We arrive at these conclusions both by making simple analytical estimates (Section \ref{ch5:sec:analytical_calculation}) and by performing  population synthesis simulations (Section \ref{ch5:sec:numerical_simulations}), with varying input parameters to account for the uncertainties.
\item There is a variety of scenarios of binary interaction that SN\,II progenitors follow (Section \ref{ch5:sec:overview_of_paths}). Almost all of them involve either mass accretion onto the progenitor star from its initially more massive companion or some type of merger. The importance of each of these channels varies for different assumptions, but we find that the overall  rate of interacting binary progenitors is significant, even for extreme assumptions (Section \ref{ch5:sec:numerical_simulations}). 
% that get ejected
\item Merger scenarios for SN\,II progenitors are expected to be common (12\%-44\%) and not an exception. At the moment of merging the two stars can either be at the beginning of their life (MS+MS mergers), or one may have evolved past its main sequence phase (postMS+MS mergers). Alternatively, merging can be initiated by a reverse mass transfer phase of the initially less massive secondary towards the stripped remnant core of the primary (Section \ref{ch5:sec:numerical_simulations}). 
\item The fraction of SN\,II progenitors that originate from secondary stars that accrete mass is also significant (3\%-18\%). These stars are ejected from the binary system due to the prior explosion of the primary, the former producing a SN\,II possibly far away from their birth location (Section \ref{ch5:sec:numerical_simulations}).

\item Although Type II-P progenitors as usually assumed to have originated from single-star evolution, our results imply that a significant fraction of them probably had mass exchange before exploding (Section \ref{ch5:sec:global_comparison}). For our standard assumptions this is the case for about one third of all Type II-P, although this is subject both to model and empirical uncertainties. 
\item The variety of the possible binary evolutionary channels may give rise to a diverse range of observed properties of SNe\,II and may account for at least part of the SN\,II subclasses, including Type IIn, Type II-L or Type II-peculiar from BSG progenitors (Section \ref{ch5:sec:diversity_type}). 

\item We find that a stripped-envelope SN is expected for around every two SNe\,II (Section \ref{ch5:sec:numerical_simulations}), which is roughly consistent with the observed SN rates. 
\item We expect a fraction of SN\,II progenitors to have drifted away from their birth place both because of their long delay times and because part of them will be ejected by their binary systems due to a prior explosion of the companion. This can lead to them being associated with older populations than expected from single stellar evolution (Section \ref{ch5:sec:environment}). No companions are predicted at the moment of explosion, save for companion stars in wide orbits that have not interacted with the progenitor (Section \ref{ch5:sec:no_companions}). 
\end{itemize}

\vspace{1em}
Our results show that a significant fraction of  SNe\,II are expected to follow binary evolutionary channels. It would be interesting to study in more depth the impact of mass exchange due to binarity on the stellar structure of SN progenitors and use binary models as input to Type II SN simulations of the explosion mechanism \citep[e.g.,][]{Burrows+2006,Janka2012,OConnor+2013}, the nucleosynthesis processes  \citep[e.g.,][]{Woosley+2002, Woosley+2007a,Sukhbold+2016}, or the light curve and the spectral evolution of the SN \citep[e.g.,][]{Bersten+2011,Dessart+2013}, rather than the single star evolution models used so far. \citet{Justham+2014}, \citet{Menon+2017} and \citet{Urushibata+2018} are studies that have explored these possibilities.

%\acknowledgments
\begin{acknowledgements}

We thank the referee for their constructive report that helped improve the manuscript. 
We are grateful to Rob Izzard for providing the population synthesis code, $\tt{binary\_c}$, used in this study and for his technical support. We are also grateful for the very useful discussions with Stephen Smartt, Maria Drout, Tassos Fragos, Rubina Kotak, Maryam Modjaz and Onno Pols. 
EZ was supported by the Netherlands Research School for Astronomy (NOVA) during most of the period of this work. 
SdM, SJ, MR and YG acknowledge funding by the European Union's Horizon 2020 research and innovation programme from the European Research Council (ERC) (Project BinCosmos, Grant agreement No. 715063), and by the Netherlands Organisation for Scientific Research (NWO) as part of the Vidi research program BinWaves with project number 639.042.728. RF is supported by an NWO top grant with project number 614.001.501 (Pi SdM). 

\end{acknowledgements}

%\input{paper_typeII_for_arxiv}
%\bibliography{my_bib_paper1_AV(fromthesis),my_bib_paper3__SJ_addition,manos_additions_paper3(fromthesis),manos_bib_paper3(fromthesis)}

\begin{thebibliography}{172}
\expandafter\ifx\csname natexlab\endcsname\relax\def\natexlab#1{#1}\fi

\bibitem[{{Almeida} {et~al.}(2017){Almeida}, {Sana}, {Taylor}, {Barb{\'a}},
  {Bonanos}, {Crowther}, {Damineli}, {de Koter}, {de Mink}, {Evans}, {Gieles},
  {Grin}, {H{\'e}nault-Brunet}, {Langer}, {Lennon}, {Lockwood}, {Ma{\'{\i}}z
  Apell{\'a}niz}, {Moffat}, {Neijssel}, {Norman}, {Ram{\'{\i}}rez-Agudelo},
  {Richardson}, {Schootemeijer}, {Shenar}, {Soszy{\'n}ski}, {Tramper}, \&
  {Vink}}]{Almeida+2017}
{Almeida}, L.~A., {Sana}, H., {Taylor}, W., {et~al.} 2017, \aap, 598, A84

\bibitem[{{Anderson} {et~al.}(2012){Anderson}, {Habergham}, {James}, \&
  {Hamuy}}]{Anderson+2012}
{Anderson}, J.~P., {Habergham}, S.~M., {James}, P.~A., \& {Hamuy}, M. 2012,
  \mnras, 424, 1372

\bibitem[{{Arcavi}(2017)}]{Arcavi2017}
{Arcavi}, I. 2017, {Hydrogen-Rich Core-Collapse Supernovae}, ed. A.~W.
  {Alsabti} \& P.~{Murdin}, 239

\bibitem[{{Auchettl} {et~al.}(2018){Auchettl}, {Lopez}, {Badenes},
  {Ramirez-Ruiz}, {Beacom}, \& {Holland-Ashford}}]{Auchettl+2018}
{Auchettl}, K., {Lopez}, L., {Badenes}, C., {et~al.} 2018, ArXiv e-prints
  [\eprint[arXiv]{1804.10210}]

\bibitem[{{Baade} \& {Zwicky}(1934)}]{Baade+1934}
{Baade}, W. \& {Zwicky}, F. 1934, Proceedings of the National Academy of
  Science, 20, 254

\bibitem[{{Begelman} \& {Sarazin}(1986)}]{Begelman+1986}
{Begelman}, M.~C. \& {Sarazin}, C.~L. 1986, \apjl, 302, L59

\bibitem[{{Bellm}(2014)}]{Bellm2014}
{Bellm}, E. 2014, in The Third Hot-wiring the Transient Universe Workshop, ed.
  P.~R. {Wozniak}, M.~J. {Graham}, A.~A. {Mahabal}, \& R.~{Seaman}, 27--33

\bibitem[{{Bersten} {et~al.}(2011){Bersten}, {Benvenuto}, \&
  {Hamuy}}]{Bersten+2011}
{Bersten}, M.~C., {Benvenuto}, O., \& {Hamuy}, M. 2011, \apj, 729, 61

\bibitem[{{Bethe} {et~al.}(1979){Bethe}, {Brown}, {Applegate}, \&
  {Lattimer}}]{Bethe+1979}
{Bethe}, H.~A., {Brown}, G.~E., {Applegate}, J., \& {Lattimer}, J.~M. 1979,
  Nuclear Physics A, 324, 487

\bibitem[{{Blaauw}(1961)}]{Blaauw1961}
{Blaauw}, A. 1961, \bain, 15, 265

\bibitem[{{Braun} \& {Langer}(1995)}]{Braun+1995}
{Braun}, H. \& {Langer}, N. 1995, \aap, 297, 483

\bibitem[{{Burrows} {et~al.}(2006){Burrows}, {Livne}, {Dessart}, {Ott}, \&
  {Murphy}}]{Burrows+2006}
{Burrows}, A., {Livne}, E., {Dessart}, L., {Ott}, C.~D., \& {Murphy}, J. 2006,
  \apj, 640, 878

\bibitem[{{Cantiello} {et~al.}(2007){Cantiello}, {Yoon}, {Langer}, \&
  {Livio}}]{Cantiello+2007}
{Cantiello}, M., {Yoon}, S.-C., {Langer}, N., \& {Livio}, M. 2007, \aap, 465,
  L29

\bibitem[{{Cao} {et~al.}(2013){Cao}, {Kasliwal}, {Arcavi}, {Horesh}, {Hancock},
  {Valenti}, {Cenko}, {Kulkarni}, {Gal-Yam}, {Gorbikov}, {Ofek}, {Sand},
  {Yaron}, {Graham}, {Silverman}, {Wheeler}, {Marion}, {Walker}, {Mazzali},
  {Howell}, {Li}, {Kong}, {Bloom}, {Nugent}, {Surace}, {Masci}, {Carpenter},
  {Degenaar}, \& {Gelino}}]{Cao+2013}
{Cao}, Y., {Kasliwal}, M.~M., {Arcavi}, I., {et~al.} 2013, \apjl, 775, L7

\bibitem[{{Chini} {et~al.}(2012){Chini}, {Hoffmeister}, {Nasseri}, {Stahl}, \&
  {Zinnecker}}]{Chini+2012}
{Chini}, R., {Hoffmeister}, V.~H., {Nasseri}, A., {Stahl}, O., \& {Zinnecker},
  H. 2012, \mnras, 424, 1925

\bibitem[{{Claeys} {et~al.}(2011){Claeys}, {de Mink}, {Pols}, {Eldridge}, \&
  {Baes}}]{Claeys+2011}
{Claeys}, J.~S.~W., {de Mink}, S.~E., {Pols}, O.~R., {Eldridge}, J.~J., \&
  {Baes}, M. 2011, \aap, 528, A131

\bibitem[{{De Donder} \& {Vanbeveren}(2003)}]{De-Donder+2003a}
{De Donder}, E. \& {Vanbeveren}, D. 2003, \na, 8, 817

\bibitem[{{De Donder} {et~al.}(1997){De Donder}, {Vanbeveren}, \& {van
  Bever}}]{de-Donder+1997}
{De Donder}, E., {Vanbeveren}, D., \& {van Bever}, J. 1997, \aap, 318, 812

\bibitem[{{de Mink} \& {Belczynski}(2015)}]{de-Mink+2015}
{de Mink}, S.~E. \& {Belczynski}, K. 2015, \apj, 814, 58

\bibitem[{{de Mink} {et~al.}(2013){de Mink}, {Langer}, {Izzard}, {Sana}, \& {de
  Koter}}]{de-Mink+2013}
{de Mink}, S.~E., {Langer}, N., {Izzard}, R.~G., {Sana}, H., \& {de Koter}, A.
  2013, \apj, 764, 166

\bibitem[{{de Mink} {et~al.}(2007){de Mink}, {Pols}, \&
  {Hilditch}}]{de-Mink+2007}
{de Mink}, S.~E., {Pols}, O.~R., \& {Hilditch}, R.~W. 2007, \aap, 467, 1181

\bibitem[{{de Mink} {et~al.}(2014){de Mink}, {Sana}, {Langer}, {Izzard}, \&
  {Schneider}}]{de-Mink+2014}
{de Mink}, S.~E., {Sana}, H., {Langer}, N., {Izzard}, R.~G., \& {Schneider},
  F.~R.~N. 2014, \apj, 782, 7

\bibitem[{{Dessart} {et~al.}(2013){Dessart}, {Hillier}, {Waldman}, \&
  {Livne}}]{Dessart+2013}
{Dessart}, L., {Hillier}, D.~J., {Waldman}, R., \& {Livne}, E. 2013, \mnras,
  433, 1745

\bibitem[{{Dewi} \& {Tauris}(2000)}]{Dewi+2000}
{Dewi}, J.~D.~M. \& {Tauris}, T.~M. 2000, \aap, 360, 1043

\bibitem[{{Dewi} \& {Tauris}(2001)}]{Dewi+2001}
{Dewi}, J.~D.~M. \& {Tauris}, T.~M. 2001, in Astronomical Society of the
  Pacific Conference Series, Vol. 229, Evolution of Binary and Multiple Star
  Systems, ed. P.~{Podsiadlowski}, S.~{Rappaport}, A.~R. {King}, F.~{D'Antona},
  \& L.~{Burderi}, 255

\bibitem[{{D{\'{\i}}az-Rodr{\'{\i}}guez}
  {et~al.}(2018){D{\'{\i}}az-Rodr{\'{\i}}guez}, {Murphy}, {Rubin}, {Dolphin},
  {Williams}, \& {Dalcanton}}]{Diaz-Rodriguez+2018}
{D{\'{\i}}az-Rodr{\'{\i}}guez}, M., {Murphy}, J.~W., {Rubin}, D.~A., {et~al.}
  2018, \apj, 861, 92

\bibitem[{{Dray} \& {Tout}(2007)}]{Dray+2007}
{Dray}, L.~M. \& {Tout}, C.~A. 2007, \mnras, 376, 61

\bibitem[{{Drout} {et~al.}(2011){Drout}, {Soderberg}, {Gal-Yam}, {Cenko},
  {Fox}, {Leonard}, {Sand}, {Moon}, {Arcavi}, \& {Green}}]{Drout+2011}
{Drout}, M.~R., {Soderberg}, A.~M., {Gal-Yam}, A., {et~al.} 2011, \apj, 741, 97

\bibitem[{{Duch{\^e}ne} \& {Kraus}(2013)}]{Duchene+2013}
{Duch{\^e}ne}, G. \& {Kraus}, A. 2013, \araa, 51, 269

\bibitem[{{Dunstall} {et~al.}(2015){Dunstall}, {Dufton}, {Sana}, {Evans},
  {Howarth}, {Sim{\'o}n-D{\'{\i}}az}, {de Mink}, {Langer}, {Ma{\'{\i}}z
  Apell{\'a}niz}, \& {Taylor}}]{Dunstall+2015}
{Dunstall}, P.~R., {Dufton}, P.~L., {Sana}, H., {et~al.} 2015, \aap, 580, A93

\bibitem[{{Eggleton} \& {Tokovinin}(2008)}]{Eggleton+2008}
{Eggleton}, P.~P. \& {Tokovinin}, A.~A. 2008, \mnras, 389, 869

\bibitem[{{Eldridge} {et~al.}(2013){Eldridge}, {Fraser}, {Smartt}, {Maund}, \&
  {Crockett}}]{Eldridge+2013}
{Eldridge}, J.~J., {Fraser}, M., {Smartt}, S.~J., {Maund}, J.~R., \&
  {Crockett}, R.~M. 2013, \mnras, 436, 774

\bibitem[{{Eldridge} {et~al.}(2008){Eldridge}, {Izzard}, \&
  {Tout}}]{Eldridge+2008}
{Eldridge}, J.~J., {Izzard}, R.~G., \& {Tout}, C.~A. 2008, \mnras, 384, 1109

\bibitem[{{Eldridge} {et~al.}(2011){Eldridge}, {Langer}, \&
  {Tout}}]{Eldridge+2011}
{Eldridge}, J.~J., {Langer}, N., \& {Tout}, C.~A. 2011, \mnras, 414, 3501

\bibitem[{{Eldridge} {et~al.}(2018){Eldridge}, {Xiao}, {Stanway}, {Rodrigues},
  \& {Guo}}]{Eldridge+2018}
{Eldridge}, J.~J., {Xiao}, L., {Stanway}, E.~R., {Rodrigues}, N., \& {Guo},
  N.-Y. 2018, \pasa, 35 [\eprint[arXiv]{1811.00282}]

\bibitem[{{Ensman} \& {Woosley}(1988)}]{Ensman+1988}
{Ensman}, L.~M. \& {Woosley}, S.~E. 1988, \apj, 333, 754

\bibitem[{{Filippenko}(1997)}]{Filippenko1997}
{Filippenko}, A.~V. 1997, \araa, 35, 309

\bibitem[{{Fuller}(2017)}]{Fuller2017}
{Fuller}, J. 2017, \mnras, 470, 1642

\bibitem[{{Gaburov} {et~al.}(2008){Gaburov}, {Lombardi}, \& {Portegies
  Zwart}}]{Gaburov+2008}
{Gaburov}, E., {Lombardi}, J.~C., \& {Portegies Zwart}, S. 2008, \mnras, 383,
  L5

\bibitem[{{Gal-Yam}(2017)}]{Gal-Yam2017}
{Gal-Yam}, A. 2017, {Observational and Physical Classification of Supernovae},
  ed. A.~W. {Alsabti} \& P.~{Murdin}, 195

\bibitem[{{Gal-Yam} {et~al.}(2007){Gal-Yam}, {Leonard}, {Fox}, {Cenko},
  {Soderberg}, {Moon}, {Sand}, {Caltech Core Collapse Program}, {Li},
  {Filippenko}, {Aldering}, \& {Copin}}]{Gal-Yam+2007}
{Gal-Yam}, A., {Leonard}, D.~C., {Fox}, D.~B., {et~al.} 2007, \apj, 656, 372

\bibitem[{{Gaskell} {et~al.}(1986){Gaskell}, {Cappellaro}, {Dinerstein},
  {Garnett}, {Harkness}, \& {Wheeler}}]{Gaskell+1986}
{Gaskell}, C.~M., {Cappellaro}, E., {Dinerstein}, H.~L., {et~al.} 1986, \apjl,
  306, L77

\bibitem[{{Georgy} {et~al.}(2012){Georgy}, {Ekstr{\"o}m}, {Meynet}, {Massey},
  {Levesque}, {Hirschi}, {Eggenberger}, \& {Maeder}}]{Georgy+2012}
{Georgy}, C., {Ekstr{\"o}m}, S., {Meynet}, G., {et~al.} 2012, \aap, 542, A29

\bibitem[{{Glebbeek} {et~al.}(2013){Glebbeek}, {Gaburov}, {Portegies Zwart}, \&
  {Pols}}]{Glebbeek+2013}
{Glebbeek}, E., {Gaburov}, E., {Portegies Zwart}, S., \& {Pols}, O.~R. 2013,
  \mnras, 434, 3497

\bibitem[{{G{\"o}tberg} {et~al.}(2017){G{\"o}tberg}, {de Mink}, \&
  {Groh}}]{Gotberg+2017}
{G{\"o}tberg}, Y., {de Mink}, S.~E., \& {Groh}, J.~H. 2017, \aap, 608, A11

\bibitem[{{Graur} {et~al.}(2017{\natexlab{a}}){Graur}, {Bianco}, {Huang},
  {Modjaz}, {Shivvers}, {Filippenko}, {Li}, \& {Eldridge}}]{Graur+2017}
{Graur}, O., {Bianco}, F.~B., {Huang}, S., {et~al.} 2017{\natexlab{a}}, \apj,
  837, 120

\bibitem[{{Graur} {et~al.}(2017{\natexlab{b}}){Graur}, {Bianco}, {Modjaz},
  {Shivvers}, {Filippenko}, {Li}, \& {Smith}}]{Graur+2017a}
{Graur}, O., {Bianco}, F.~B., {Modjaz}, M., {et~al.} 2017{\natexlab{b}}, \apj,
  837, 121

\bibitem[{{Groh} {et~al.}(2013{\natexlab{a}}){Groh}, {Georgy}, \&
  {Ekstr{\"o}m}}]{Groh+2013b}
{Groh}, J.~H., {Georgy}, C., \& {Ekstr{\"o}m}, S. 2013{\natexlab{a}}, \aap,
  558, L1

\bibitem[{{Groh} {et~al.}(2013{\natexlab{b}}){Groh}, {Meynet}, {Georgy}, \&
  {Ekstr{\"o}m}}]{Groh+2013a}
{Groh}, J.~H., {Meynet}, G., {Georgy}, C., \& {Ekstr{\"o}m}, S.
  2013{\natexlab{b}}, \aap, 558, A131

\bibitem[{{Grunhut} {et~al.}(2013){Grunhut}, {Wade}, {Leutenegger}, {Petit},
  {Rauw}, {Neiner}, {Martins}, {Cohen}, {Gagn{\'e}}, {Ignace}, {Mathis}, {de
  Mink}, {Moffat}, {Owocki}, {Shultz}, {Sundqvist}, \& {MiMeS
  Collaboration}}]{Grunhut+2013}
{Grunhut}, J.~H., {Wade}, G.~A., {Leutenegger}, M., {et~al.} 2013, \mnras, 428,
  1686

\bibitem[{{Habergham} {et~al.}(2014){Habergham}, {Anderson}, {James}, \&
  {Lyman}}]{Habergham+2014}
{Habergham}, S.~M., {Anderson}, J.~P., {James}, P.~A., \& {Lyman}, J.~D. 2014,
  \mnras, 441, 2230

\bibitem[{{Hamann} {et~al.}(1995){Hamann}, {Koesterke}, \&
  {Wessolowski}}]{Hamann+1995}
{Hamann}, W.-R., {Koesterke}, L., \& {Wessolowski}, U. 1995, \aap, 299, 151

\bibitem[{{Hamuy} {et~al.}(2003){Hamuy}, {Phillips}, {Suntzeff}, {Maza},
  {Gonz{\'a}lez}, {Roth}, {Krisciunas}, {Morrell}, {Green}, {Persson}, \&
  {McCarthy}}]{Hamuy+2003}
{Hamuy}, M., {Phillips}, M.~M., {Suntzeff}, N.~B., {et~al.} 2003, \nat, 424,
  651

\bibitem[{{Han} \& {Podsiadlowski}(2006)}]{Han+PhP2006}
{Han}, Z. \& {Podsiadlowski}, P. 2006, \mnras, 368, 1095

\bibitem[{{Heger} {et~al.}(2003){Heger}, {Fryer}, {Woosley}, {Langer}, \&
  {Hartmann}}]{Heger+2003}
{Heger}, A., {Fryer}, C.~L., {Woosley}, S.~E., {Langer}, N., \& {Hartmann},
  D.~H. 2003, \apj, 591, 288

\bibitem[{{Hellings}(1983)}]{Hellings1983}
{Hellings}, P. 1983, \apss, 96, 37

\bibitem[{{Hellings}(1984)}]{Hellings1984}
{Hellings}, P. 1984, \apss, 104, 83

\bibitem[{{Hobbs} {et~al.}(2005){Hobbs}, {Lorimer}, {Lyne}, \&
  {Kramer}}]{Hobbs+2005}
{Hobbs}, G., {Lorimer}, D.~R., {Lyne}, A.~G., \& {Kramer}, M. 2005, \mnras,
  360, 974

\bibitem[{{Humphreys} {et~al.}(2016){Humphreys}, {Weis}, {Davidson}, \&
  {Gordon}}]{Humphreys+2016}
{Humphreys}, R.~M., {Weis}, K., {Davidson}, K., \& {Gordon}, M.~S. 2016, \apj,
  825, 64

\bibitem[{{Hurley} {et~al.}(2000){Hurley}, {Pols}, \& {Tout}}]{Hurley+2000}
{Hurley}, J.~R., {Pols}, O.~R., \& {Tout}, C.~A. 2000, \mnras, 315, 543

\bibitem[{{Hurley} {et~al.}(2002){Hurley}, {Tout}, \& {Pols}}]{Hurley+2002}
{Hurley}, J.~R., {Tout}, C.~A., \& {Pols}, O.~R. 2002, \mnras, 329, 897

\bibitem[{{Hut}(1980)}]{Hut1980}
{Hut}, P. 1980, \aap, 92, 167

\bibitem[{{Hut}(1981)}]{Hut1981}
{Hut}, P. 1981, \aap, 99, 126

\bibitem[{{Ivanova} {et~al.}(2013){Ivanova}, {Justham}, {Chen}, {De Marco},
  {Fryer}, {Gaburov}, {Ge}, {Glebbeek}, {Han}, {Li}, {Lu}, {Marsh},
  {Podsiadlowski}, {Potter}, {Soker}, {Taam}, {Tauris}, {van den Heuvel}, \&
  {Webbink}}]{Ivanova+2013}
{Ivanova}, N., {Justham}, S., {Chen}, X., {et~al.} 2013, \aapr, 21, 59

\bibitem[{{Izzard} {et~al.}(2006){Izzard}, {Dray}, {Karakas}, {Lugaro}, \&
  {Tout}}]{Izzard+2006}
{Izzard}, R.~G., {Dray}, L.~M., {Karakas}, A.~I., {Lugaro}, M., \& {Tout},
  C.~A. 2006, \aap, 460, 565

\bibitem[{{Izzard} {et~al.}(2009){Izzard}, {Glebbeek}, {Stancliffe}, \&
  {Pols}}]{Izzard+2009}
{Izzard}, R.~G., {Glebbeek}, E., {Stancliffe}, R.~J., \& {Pols}, O.~R. 2009,
  \aap, 508, 1359

\bibitem[{{Izzard} {et~al.}(2004){Izzard}, {Tout}, {Karakas}, \&
  {Pols}}]{Izzard+2004}
{Izzard}, R.~G., {Tout}, C.~A., {Karakas}, A.~I., \& {Pols}, O.~R. 2004,
  \mnras, 350, 407

\bibitem[{{Janka}(2012)}]{Janka2012}
{Janka}, H.-T. 2012, Annual Review of Nuclear and Particle Science, 62, 407

\bibitem[{{Jennings} {et~al.}(2014){Jennings}, {Williams}, {Murphy},
  {Dalcanton}, {Gilbert}, {Dolphin}, {Weisz}, \& {Fouesneau}}]{Jennings+2014}
{Jennings}, Z.~G., {Williams}, B.~F., {Murphy}, J.~W., {et~al.} 2014, \apj,
  795, 170

\bibitem[{{Justham} {et~al.}(2014){Justham}, {Podsiadlowski}, \&
  {Vink}}]{Justham+2014}
{Justham}, S., {Podsiadlowski}, P., \& {Vink}, J.~S. 2014, \apj, 796, 121

\bibitem[{{Kaiser} {et~al.}(2010){Kaiser}, {Burgett}, {Chambers}, {Denneau},
  {Heasley}, {Jedicke}, {Magnier}, {Morgan}, {Onaka}, \& {Tonry}}]{Kaiser+2010}
{Kaiser}, N., {Burgett}, W., {Chambers}, K., {et~al.} 2010, in \procspie, Vol.
  7733, Ground-based and Airborne Telescopes III, 77330E

\bibitem[{{Khazov} {et~al.}(2016){Khazov}, {Yaron}, {Gal-Yam}, {Manulis},
  {Rubin}, {Kulkarni}, {Arcavi}, {Kasliwal}, {Ofek}, {Cao}, {Perley},
  {Sollerman}, {Horesh}, {Sullivan}, {Filippenko}, {Nugent}, {Howell}, {Cenko},
  {Silverman}, {Ebeling}, {Taddia}, {Johansson}, {Laher}, {Surace},
  {Rebbapragada}, {Wozniak}, \& {Matheson}}]{Khazov+2016}
{Khazov}, D., {Yaron}, O., {Gal-Yam}, A., {et~al.} 2016, \apj, 818, 3

\bibitem[{{Kiminki} \& {Kobulnicky}(2012)}]{Kiminki+2012}
{Kiminki}, D.~C. \& {Kobulnicky}, H.~A. 2012, \apj, 751, 4

\bibitem[{{Kippenhahn} \& {Weigert}(1967)}]{Kippenhahn+1967}
{Kippenhahn}, R. \& {Weigert}, A. 1967, \zap, 65, 251

\bibitem[{{Kobulnicky} \& {Fryer}(2007)}]{Kobulnicky+2007}
{Kobulnicky}, H.~A. \& {Fryer}, C.~L. 2007, \apj, 670, 747

\bibitem[{{Kochanek}(2017)}]{Kochanek2017}
{Kochanek}, C.~S. 2017, \mnras, 471, 3283

\bibitem[{{Kochanek}(2018)}]{Kochanek2018}
{Kochanek}, C.~S. 2018, \mnras, 473, 1633

\bibitem[{{Kotak} {et~al.}(2004){Kotak}, {Meikle}, {Adamson}, \&
  {Leggett}}]{Kotak+2004}
{Kotak}, R., {Meikle}, W.~P.~S., {Adamson}, A., \& {Leggett}, S.~K. 2004,
  \mnras, 354, L13

\bibitem[{{Kroupa}(2001)}]{Kroupa2001}
{Kroupa}, P. 2001, \mnras, 322, 231

\bibitem[{{Kuncarayakti} {et~al.}(2018){Kuncarayakti}, {Anderson}, {Galbany},
  {Maeda}, {Hamuy}, {Aldering}, {Arimoto}, {Doi}, {Morokuma}, \&
  {Usuda}}]{Kuncarayakti+2018}
{Kuncarayakti}, H., {Anderson}, J.~P., {Galbany}, L., {et~al.} 2018, \aap, 613,
  A35

\bibitem[{{Langer}(1991)}]{Langer1991a}
{Langer}, N. 1991, \aap, 243, 155

\bibitem[{{Li} {et~al.}(2011){Li}, {Leaman}, {Chornock}, {Filippenko},
  {Poznanski}, {Ganeshalingam}, {Wang}, {Modjaz}, {Jha}, {Foley}, \&
  {Smith}}]{Li+2011}
{Li}, W., {Leaman}, J., {Chornock}, R., {et~al.} 2011, \mnras, 412, 1441

\bibitem[{{Li} {et~al.}(2006){Li}, {Van Dyk}, {Filippenko}, {Cuillandre},
  {Jha}, {Bloom}, {Riess}, \& {Livio}}]{Li+2006}
{Li}, W., {Van Dyk}, S.~D., {Filippenko}, A.~V., {et~al.} 2006, \apj, 641, 1060

\bibitem[{{Lombardi} {et~al.}(1995){Lombardi}, {Rasio}, \&
  {Shapiro}}]{Lombardi+1995}
{Lombardi}, Jr., J.~C., {Rasio}, F.~A., \& {Shapiro}, S.~L. 1995, \apjl, 445,
  L117

\bibitem[{{Lombardi} {et~al.}(1996){Lombardi}, {Rasio}, \&
  {Shapiro}}]{Lombardi+1996}
{Lombardi}, Jr., J.~C., {Rasio}, F.~A., \& {Shapiro}, S.~L. 1996, \apj, 468,
  797

\bibitem[{{LSST Science Collaboration} {et~al.}(2009){LSST Science
  Collaboration}, {Abell}, {Allison}, {Anderson}, {Andrew}, {Angel}, {Armus},
  {Arnett}, {Asztalos}, {Axelrod}, {Bailey}, {Ballantyne}, {Bankert},
  {Barkhouse}, {Barr}, {Barrientos}, {Barth}, {Bartlett}, {Becker}, {Becla},
  {Beers}, {Bernstein}, {Biswas}, {Blanton}, {Bloom}, {Bochanski}, {Boeshaar},
  {Borne}, {Bradac}, {Brandt}, {Bridge}, {Brown}, {Brunner}, {Bullock},
  {Burgasser}, {Burge}, {Burke}, {Cargile}, {Chandrasekharan}, {Chartas},
  {Chesley}, {Chu}, {Cinabro}, {Claire}, {Claver}, {Clowe}, {Connolly}, {Cook},
  {Cooke}, {Cooray}, {Covey}, {Culliton}, {de Jong}, {de Vries}, {Debattista},
  {Delgado}, {Dell'Antonio}, {Dhital}, {Di Stefano}, {Dickinson}, {Dilday},
  {Djorgovski}, {Dobler}, {Donalek}, {Dubois-Felsmann}, {Durech},
  {Eliasdottir}, {Eracleous}, {Eyer}, {Falco}, {Fan}, {Fassnacht}, {Ferguson},
  {Fernandez}, {Fields}, {Finkbeiner}, {Figueroa}, {Fox}, {Francke}, {Frank},
  {Frieman}, {Fromenteau}, {Furqan}, {Galaz}, {Gal-Yam}, {Garnavich},
  {Gawiser}, {Geary}, {Gee}, {Gibson}, {Gilmore}, {Grace}, {Green}, {Gressler},
  {Grillmair}, {Habib}, {Haggerty}, {Hamuy}, {Harris}, {Hawley}, {Heavens},
  {Hebb}, {Henry}, {Hileman}, {Hilton}, {Hoadley}, {Holberg}, {Holman},
  {Howell}, {Infante}, {Ivezic}, {Jacoby}, {Jain}, {R}, {Jedicke}, {Jee},
  {Garrett Jernigan}, {Jha}, {Johnston}, {Jones}, {Juric}, {Kaasalainen},
  {Styliani}, {Kafka}, {Kahn}, {Kaib}, {Kalirai}, {Kantor}, {Kasliwal},
  {Keeton}, {Kessler}, {Knezevic}, {Kowalski}, {Krabbendam}, {Krughoff},
  {Kulkarni}, {Kuhlman}, {Lacy}, {Lepine}, {Liang}, {Lien}, {Lira}, {Long},
  {Lorenz}, {Lotz}, {Lupton}, {Lutz}, {Macri}, {Mahabal}, {Mandelbaum},
  {Marshall}, {May}, {McGehee}, {Meadows}, {Meert}, {Milani}, {Miller},
  {Miller}, {Mills}, {Minniti}, {Monet}, {Mukadam}, {Nakar}, {Neill}, {Newman},
  {Nikolaev}, {Nordby}, {O'Connor}, {Oguri}, {Oliver}, {Olivier}, {Olsen},
  {Olsen}, {Olszewski}, {Oluseyi}, {Padilla}, {Parker}, {Pepper}, {Peterson},
  {Petry}, {Pinto}, {Pizagno}, {Popescu}, {Prsa}, {Radcka}, {Raddick},
  {Rasmussen}, {Rau}, {Rho}, {Rhoads}, {Richards}, {Ridgway}, {Robertson},
  {Roskar}, {Saha}, {Sarajedini}, {Scannapieco}, {Schalk}, {Schindler},
  {Schmidt}, {Schmidt}, {Schneider}, {Schumacher}, {Scranton}, {Sebag},
  {Seppala}, {Shemmer}, {Simon}, {Sivertz}, {Smith}, {Allyn Smith}, {Smith},
  {Spitz}, {Stanford}, {Stassun}, {Strader}, {Strauss}, {Stubbs}, {Sweeney},
  {Szalay}, {Szkody}, {Takada}, {Thorman}, {Trilling}, {Trimble}, {Tyson}, {Van
  Berg}, {Vanden Berk}, {VanderPlas}, {Verde}, {Vrsnak}, {Walkowicz},
  {Wandelt}, {Wang}, {Wang}, {Warner}, {Wechsler}, {West}, {Wiecha},
  {Williams}, {Willman}, {Wittman}, {Wolff}, {Wood-Vasey}, {Wozniak}, {Young},
  {Zentner}, \& {Zhan}}]{LSST-Science-Collaboration+2009}
{LSST Science Collaboration}, {Abell}, P.~A., {Allison}, J., {et~al.} 2009,
  ArXiv e-prints [\eprint[arXiv]{0912.0201}]

\bibitem[{{Lyman} {et~al.}(2016){Lyman}, {Bersier}, {James}, {Mazzali},
  {Eldridge}, {Fraser}, \& {Pian}}]{Lyman+2016}
{Lyman}, J.~D., {Bersier}, D., {James}, P.~A., {et~al.} 2016, \mnras, 457, 328

\bibitem[{{Maeder} \& {Meynet}(2000)}]{Maeder+2000a}
{Maeder}, A. \& {Meynet}, G. 2000, \araa, 38, 143

\bibitem[{{Mattila} {et~al.}(2008){Mattila}, {Smartt}, {Eldridge}, {Maund},
  {Crockett}, \& {Danziger}}]{Mattila+2008}
{Mattila}, S., {Smartt}, S.~J., {Eldridge}, J.~J., {et~al.} 2008, \apjl, 688,
  L91

\bibitem[{{Maund}(2017)}]{Maund2017}
{Maund}, J.~R. 2017, \mnras, 469, 2202

\bibitem[{{Maund} \& {Smartt}(2005)}]{Maund+2005a}
{Maund}, J.~R. \& {Smartt}, S.~J. 2005, \mnras, 360, 288

\bibitem[{{Maund} {et~al.}(2005){Maund}, {Smartt}, \& {Danziger}}]{Maund+2005}
{Maund}, J.~R., {Smartt}, S.~J., \& {Danziger}, I.~J. 2005, \mnras, 364, L33

\bibitem[{{Mcley} \& {Soker}(2014)}]{Mcley+2014}
{Mcley}, L. \& {Soker}, N. 2014, \mnras, 445, 2492

\bibitem[{{Menon} \& {Heger}(2017)}]{Menon+2017}
{Menon}, A. \& {Heger}, A. 2017, \mnras, 469, 4649

\bibitem[{{Mestel}(1957)}]{Mestel1957}
{Mestel}, L. 1957, \apj, 126, 550

\bibitem[{{Mestel} \& {Moss}(1986)}]{Mestel+1986}
{Mestel}, L. \& {Moss}, D.~L. 1986, \mnras, 221, 25

\bibitem[{{Moe} \& {Di Stefano}(2017)}]{Moe+2017}
{Moe}, M. \& {Di Stefano}, R. 2017, \apjs, 230, 15

\bibitem[{{Morozova} {et~al.}(2016){Morozova}, {Piro}, {Renzo}, \&
  {Ott}}]{Morozova+2016}
{Morozova}, V., {Piro}, A.~L., {Renzo}, M., \& {Ott}, C.~D. 2016, \apj, 829,
  109

\bibitem[{{Morozova} {et~al.}(2017){Morozova}, {Piro}, \&
  {Valenti}}]{Morozova+2017}
{Morozova}, V., {Piro}, A.~L., \& {Valenti}, S. 2017, \apj, 838, 28

\bibitem[{{Morris} \& {Podsiadlowski}(2007)}]{Morris+2007}
{Morris}, T. \& {Podsiadlowski}, P. 2007, Science, 315, 1103

\bibitem[{{Murphy} {et~al.}(2011){Murphy}, {Jennings}, {Williams}, {Dalcanton},
  \& {Dolphin}}]{Murphy+2011}
{Murphy}, J.~W., {Jennings}, Z.~G., {Williams}, B., {Dalcanton}, J.~J., \&
  {Dolphin}, A.~E. 2011, \apjl, 742, L4

\bibitem[{{Nelson} \& {Eggleton}(2001)}]{Nelson+2001}
{Nelson}, C.~A. \& {Eggleton}, P.~P. 2001, \apj, 552, 664

\bibitem[{{Nieuwenhuijzen} \& {de Jager}(1990)}]{Nieuwenhuijzen+1990}
{Nieuwenhuijzen}, H. \& {de Jager}, C. 1990, \aap, 231, 134

\bibitem[{{Nomoto} {et~al.}(1996){Nomoto}, {Iwamoto}, {Suzuki}, {Pols},
  {Yamaoka}, {Hashimoto}, {Hoflich}, \& {van den Heuvel}}]{Nomoto+1996}
{Nomoto}, K., {Iwamoto}, K., {Suzuki}, T., {et~al.} 1996, in IAU Symposium,
  Vol. 165, Compact Stars in Binaries, ed. J.~{van Paradijs}, E.~P.~J. {van den
  Heuvel}, \& E.~{Kuulkers}, 119

\bibitem[{{O'Connor} \& {Ott}(2011)}]{OConnor+2011}
{O'Connor}, E. \& {Ott}, C.~D. 2011, \apj, 730, 70

\bibitem[{{O'Connor} \& {Ott}(2013)}]{OConnor+2013}
{O'Connor}, E. \& {Ott}, C.~D. 2013, \apj, 762, 126

\bibitem[{{Packet}(1981)}]{Packet1981}
{Packet}, W. 1981, \aap, 102, 17

\bibitem[{{Pian} \& {Mazzali}(2017)}]{Pian+2017}
{Pian}, E. \& {Mazzali}, P.~A. 2017, {Hydrogen-Poor Core-Collapse Supernovae},
  ed. A.~W. {Alsabti} \& P.~{Murdin}, 277

\bibitem[{{Podsiadlowski}(1992)}]{Podsiadlowski1992}
{Podsiadlowski}, P. 1992, \pasp, 104, 717

\bibitem[{{Podsiadlowski} \& {Joss}(1989)}]{Podsiadlowski+1989}
{Podsiadlowski}, P. \& {Joss}, P.~C. 1989, \nat, 338, 401

\bibitem[{{Podsiadlowski} {et~al.}(1992){Podsiadlowski}, {Joss}, \&
  {Hsu}}]{Podsiadlowski+1992}
{Podsiadlowski}, P., {Joss}, P.~C., \& {Hsu}, J.~J.~L. 1992, \apj, 391, 246

\bibitem[{{Podsiadlowski} {et~al.}(1990){Podsiadlowski}, {Joss}, \&
  {Rappaport}}]{Podsiadlowski+1990}
{Podsiadlowski}, P., {Joss}, P.~C., \& {Rappaport}, S. 1990, \aap, 227, L9

\bibitem[{{Pols}(1994)}]{Pols1994}
{Pols}, O.~R. 1994, \aap, 290, 119

\bibitem[{{Pols} {et~al.}(1998){Pols}, {Schr\"oder}, {Hurley}, {Tout}, \&
  {Eggleton}}]{Pols+1998}
{Pols}, O.~R., {Schr\"oder}, K.-P., {Hurley}, J.~R., {Tout}, C.~A., \&
  {Eggleton}, P.~P. 1998, \mnras, 298, 525

\bibitem[{{Popov}(1993)}]{Popov1993}
{Popov}, D.~V. 1993, \apj, 414, 712

\bibitem[{{Rappaport} \& {van den Heuvel}(1982)}]{Rappaport+1982}
{Rappaport}, S. \& {van den Heuvel}, E.~P.~J. 1982, in IAU Symposium, Vol.~98,
  Be Stars, ed. {M.~Jaschek \& H.-G.~Groth}, 327

\bibitem[{{Renzo} {et~al.}(2019){Renzo}, {Zapartas}, {de Mink}, {G{\"o}tberg},
  {Justham}, {Farmer}, {Izzard}, {Toonen}, \& {Sana}}]{Renzo+2019}
{Renzo}, M., {Zapartas}, E., {de Mink}, S.~E., {et~al.} 2019, \aap, 624, A66

\bibitem[{{Sabach} \& {Soker}(2014)}]{Sabach+2014}
{Sabach}, E. \& {Soker}, N. 2014, \mnras, 439, 954

\bibitem[{{Sana} {et~al.}(2013){Sana}, {de Koter}, {de Mink}, {Dunstall},
  {Evans}, {H{\'e}nault-Brunet}, {Ma{\'{\i}}z Apell{\'a}niz},
  {Ram{\'{\i}}rez-Agudelo}, {Taylor}, {Walborn}, {Clark}, {Crowther},
  {Herrero}, {Gieles}, {Langer}, {Lennon}, \& {Vink}}]{Sana+2013}
{Sana}, H., {de Koter}, A., {de Mink}, S.~E., {et~al.} 2013, \aap, 550, A107

\bibitem[{Sana {et~al.}(2012)Sana, de~Mink, de~Koter, Langer, Evans, Gieles,
  Gosset, Izzard, Le~Bouquin, \& Schneider}]{Sana+2012}
Sana, H., de~Mink, S.~E., de~Koter, A., {et~al.} 2012, Science, 337, 444

\bibitem[{{Sana} {et~al.}(2014){Sana}, {Le Bouquin}, {Lacour}, {Berger},
  {Duvert}, {Gauchet}, {Norris}, {Olofsson}, {Pickel}, {Zins}, {Absil}, {de
  Koter}, {Kratter}, {Schnurr}, \& {Zinnecker}}]{Sana+2014}
{Sana}, H., {Le Bouquin}, J.-B., {Lacour}, S., {et~al.} 2014, \apjs, 215, 15

\bibitem[{{Schneider} {et~al.}(2015){Schneider}, {Izzard}, {Langer}, \& {de
  Mink}}]{Schneider+2015}
{Schneider}, F.~R.~N., {Izzard}, R.~G., {Langer}, N., \& {de Mink}, S.~E. 2015,
  \apj, 805, 20

\bibitem[{{Schneider} {et~al.}(2016){Schneider}, {Podsiadlowski}, {Langer},
  {Castro}, \& {Fossati}}]{Schneider+2016}
{Schneider}, F.~R.~N., {Podsiadlowski}, P., {Langer}, N., {Castro}, N., \&
  {Fossati}, L. 2016, \mnras, 457, 2355

\bibitem[{{Schootemeijer} {et~al.}(2018){Schootemeijer}, {G{\"o}tberg}, {Mink},
  {Gies}, \& {Zapartas}}]{Schootemeijer+2018}
{Schootemeijer}, A., {G{\"o}tberg}, Y., {Mink}, S.~E.~d., {Gies}, D., \&
  {Zapartas}, E. 2018, \aap, 615, A30

\bibitem[{{Shappee} {et~al.}(2014){Shappee}, {Prieto}, {Grupe}, {Kochanek},
  {Stanek}, {De Rosa}, {Mathur}, {Zu}, {Peterson}, {Pogge}, {Komossa}, {Im},
  {Jencson}, {Holoien}, {Basu}, {Beacom}, {Szczygie{\l}}, {Brimacombe},
  {Adams}, {Campillay}, {Choi}, {Contreras}, {Dietrich}, {Dubberley},
  {Elphick}, {Foale}, {Giustini}, {Gonzalez}, {Hawkins}, {Howell}, {Hsiao},
  {Koss}, {Leighly}, {Morrell}, {Mudd}, {Mullins}, {Nugent}, {Parrent},
  {Phillips}, {Pojmanski}, {Rosing}, {Ross}, {Sand}, {Terndrup}, {Valenti},
  {Walker}, \& {Yoon}}]{Shappee+2014}
{Shappee}, B.~J., {Prieto}, J.~L., {Grupe}, D., {et~al.} 2014, \apj, 788, 48

\bibitem[{{Shivvers} {et~al.}(2017){Shivvers}, {Modjaz}, {Zheng}, {Liu},
  {Filippenko}, {Silverman}, {Matheson}, {Pastorello}, {Graur}, {Foley},
  {Chornock}, {Smith}, {Leaman}, \& {Benetti}}]{Shivvers+2017}
{Shivvers}, I., {Modjaz}, M., {Zheng}, W., {et~al.} 2017, \pasp, 129, 054201

\bibitem[{{Silverman} {et~al.}(2013){Silverman}, {Nugent}, {Gal-Yam},
  {Sullivan}, {Howell}, {Filippenko}, {Arcavi}, {Ben-Ami}, {Bloom}, {Cenko},
  {Cao}, {Chornock}, {Clubb}, {Coil}, {Foley}, {Graham}, {Griffith}, {Horesh},
  {Kasliwal}, {Kulkarni}, {Leonard}, {Li}, {Matheson}, {Miller}, {Modjaz},
  {Ofek}, {Pan}, {Perley}, {Poznanski}, {Quimby}, {Steele}, {Sternberg}, {Xu},
  \& {Yaron}}]{Silverman+2013}
{Silverman}, J.~M., {Nugent}, P.~E., {Gal-Yam}, A., {et~al.} 2013, \apjs, 207,
  3

\bibitem[{{Smartt} {et~al.}(2009){Smartt}, {Eldridge}, {Crockett}, \&
  {Maund}}]{Smartt+2009}
{Smartt}, S.~J., {Eldridge}, J.~J., {Crockett}, R.~M., \& {Maund}, J.~R. 2009,
  \mnras, 395, 1409

\bibitem[{{Smith}(2014)}]{Smith2014}
{Smith}, N. 2014, \araa, 52, 487

\bibitem[{{Smith}(2016)}]{Smith2016}
{Smith}, N. 2016, \mnras, 461, 3353

\bibitem[{{Smith}(2017)}]{Smith2017}
{Smith}, N. 2017, {Interacting Supernovae: Types IIn and Ibn}, ed. A.~W.
  {Alsabti} \& P.~{Murdin}, 403

\bibitem[{{Smith} \& {Arnett}(2014)}]{Smith+2014}
{Smith}, N. \& {Arnett}, W.~D. 2014, \apj, 785, 82

\bibitem[{{Smith} {et~al.}(2011){Smith}, {Li}, {Filippenko}, \&
  {Chornock}}]{Smith+2011}
{Smith}, N., {Li}, W., {Filippenko}, A.~V., \& {Chornock}, R. 2011, \mnras,
  412, 1522

\bibitem[{{Smith} {et~al.}(2007){Smith}, {Li}, {Foley}, {Wheeler}, {Pooley},
  {Chornock}, {Filippenko}, {Silverman}, {Quimby}, {Bloom}, \&
  {Hansen}}]{Smith+2007}
{Smith}, N., {Li}, W., {Foley}, R.~J., {et~al.} 2007, \apj, 666, 1116

\bibitem[{{Smith} {et~al.}(2015){Smith}, {Mauerhan}, {Cenko}, {Kasliwal},
  {Silverman}, {Filippenko}, {Gal-Yam}, {Clubb}, {Graham}, {Leonard}, {Horst},
  {Williams}, {Andrews}, {Kulkarni}, {Nugent}, {Sullivan}, {Maguire}, {Xu}, \&
  {Ben-Ami}}]{Smith+2015a}
{Smith}, N., {Mauerhan}, J.~C., {Cenko}, S.~B., {et~al.} 2015, \mnras, 449,
  1876

\bibitem[{{Smith} \& {Tombleson}(2015)}]{Smith+2015}
{Smith}, N. \& {Tombleson}, R. 2015, \mnras, 447, 598

\bibitem[{{Smith} {et~al.}(2014){Smith}, {Dekany}, {Bebek}, {Bellm}, {Bui},
  {Cromer}, {Gardner}, {Hoff}, {Kaye}, {Kulkarni}, {Lambert}, {Levi}, \&
  {Reiley}}]{Smith+2014a}
{Smith}, R.~M., {Dekany}, R.~G., {Bebek}, C., {et~al.} 2014, in \procspie, Vol.
  9147, Ground-based and Airborne Instrumentation for Astronomy V, 914779

\bibitem[{{Soker} \& {Gilkis}(2018)}]{Soker+2018}
{Soker}, N. \& {Gilkis}, A. 2018, \mnras, 475, 1198

\bibitem[{{Sparks} \& {Stecher}(1974)}]{Sparks+1974}
{Sparks}, W.~M. \& {Stecher}, T.~P. 1974, \apj, 188, 149

\bibitem[{{Sukhbold} {et~al.}(2016){Sukhbold}, {Ertl}, {Woosley}, {Brown}, \&
  {Janka}}]{Sukhbold+2016}
{Sukhbold}, T., {Ertl}, T., {Woosley}, S.~E., {Brown}, J.~M., \& {Janka}, H.-T.
  2016, \apj, 821, 38

\bibitem[{{Taddia} {et~al.}(2016){Taddia}, {Sollerman}, {Fremling}, {Migotto},
  {Gal-Yam}, {Armen}, {Duggan}, {Ergon}, {Filippenko}, {Fransson},
  {Hosseinzadeh}, {Kasliwal}, {Laher}, {Leloudas}, {Leonard}, {Lunnan},
  {Masci}, {Moon}, {Silverman}, \& {Wozniak}}]{Taddia+2016}
{Taddia}, F., {Sollerman}, J., {Fremling}, C., {et~al.} 2016, \aap, 588, A5

\bibitem[{{Taddia} {et~al.}(2015){Taddia}, {Sollerman}, {Leloudas},
  {Stritzinger}, {Valenti}, {Galbany}, {Kessler}, {Schneider}, \&
  {Wheeler}}]{Taddia+2015}
{Taddia}, F., {Sollerman}, J., {Leloudas}, G., {et~al.} 2015, \aap, 574, A60

\bibitem[{{Tauris} \& {Dewi}(2001)}]{Tauris+2001}
{Tauris}, T.~M. \& {Dewi}, J.~D.~M. 2001, \aap, 369, 170

\bibitem[{{Thorne} \& {Zytkow}(1977)}]{Thorne+1977}
{Thorne}, K.~S. \& {Zytkow}, A.~N. 1977, \apj, 212, 832

\bibitem[{{Toonen} {et~al.}(2016){Toonen}, {Hamers}, \& {Portegies
  Zwart}}]{Toonen+2016}
{Toonen}, S., {Hamers}, A., \& {Portegies Zwart}, S. 2016, Computational
  Astrophysics and Cosmology, 3, 6

\bibitem[{{Tout} {et~al.}(1997){Tout}, {Aarseth}, {Pols}, \&
  {Eggleton}}]{Tout+1997}
{Tout}, C.~A., {Aarseth}, S.~J., {Pols}, O.~R., \& {Eggleton}, P.~P. 1997,
  \mnras, 291, 732

\bibitem[{{Tramper} {et~al.}(2015){Tramper}, {Straal}, {Sanyal}, {Sana}, {de
  Koter}, {Gr{\"a}fener}, {Langer}, {Vink}, {de Mink}, \&
  {Kaper}}]{Tramper+2015}
{Tramper}, F., {Straal}, S.~M., {Sanyal}, D., {et~al.} 2015, \aap, 581, A110

\bibitem[{{Ugliano} {et~al.}(2012){Ugliano}, {Janka}, {Marek}, \&
  {Arcones}}]{Ugliano+2012}
{Ugliano}, M., {Janka}, H.-T., {Marek}, A., \& {Arcones}, A. 2012, \apj, 757,
  69

\bibitem[{{Urushibata} {et~al.}(2018){Urushibata}, {Takahashi}, {Umeda}, \&
  {Yoshida}}]{Urushibata+2018}
{Urushibata}, T., {Takahashi}, K., {Umeda}, H., \& {Yoshida}, T. 2018, \mnras,
  473, L101

\bibitem[{{van den Heuvel}(1994)}]{van-den-Heuvel1994}
{van den Heuvel}, E.~P.~J. 1994, in Saas-Fee Advanced Course 22: Interacting
  Binaries, ed. S.~N. {Shore}, M.~{Livio}, E.~P.~J. {van den Heuvel},
  H.~{Nussbaumer}, \& A.~{Orr}, 263--474

\bibitem[{{Van Dyk} {et~al.}(2003){Van Dyk}, {Li}, \&
  {Filippenko}}]{Van-Dyk+2003}
{Van Dyk}, S.~D., {Li}, W., \& {Filippenko}, A.~V. 2003, \pasp, 115, 1

\bibitem[{{Van Dyk} {et~al.}(2018){Van Dyk}, {Zheng}, {Brink}, {Filippenko},
  {Milisavljevic}, {Andrews}, {Smith}, {Cignoni}, {Fox}, {Kelly}, {Adamo},
  {Yunus}, {Zhang}, \& {Kumar}}]{Van-Dyk+2018}
{Van Dyk}, S.~D., {Zheng}, W., {Brink}, T.~G., {et~al.} 2018, \apj, 860, 90

\bibitem[{{Vanbeveren} {et~al.}(2013){Vanbeveren}, {Mennekens}, {Van
  Rensbergen}, \& {De Loore}}]{Vanbeveren+2013}
{Vanbeveren}, D., {Mennekens}, N., {Van Rensbergen}, W., \& {De Loore}, C.
  2013, \aap, 552, A105

\bibitem[{{Vink} {et~al.}(2000){Vink}, {de Koter}, \& {Lamers}}]{Vink+2000}
{Vink}, J.~S., {de Koter}, A., \& {Lamers}, H.~J.~G.~L.~M. 2000, \aap, 362, 295

\bibitem[{{Vink} {et~al.}(2001){Vink}, {de Koter}, \& {Lamers}}]{Vink+2001}
{Vink}, J.~S., {de Koter}, A., \& {Lamers}, H.~J.~G.~L.~M. 2001, \aap, 369, 574

\bibitem[{{Wang} {et~al.}(2014){Wang}, {Justham}, {Liu}, {Zhang}, {Liu}, \&
  {Han}}]{WangJustham+2014}
{Wang}, B., {Justham}, S., {Liu}, Z.-W., {et~al.} 2014, \mnras, 445, 2340

\bibitem[{{Webbink}(1984)}]{Webbink1984}
{Webbink}, R.~F. 1984, \apj, 277, 355

\bibitem[{{Wellstein} {et~al.}(2001){Wellstein}, {Langer}, \&
  {Braun}}]{Wellstein+2001}
{Wellstein}, S., {Langer}, N., \& {Braun}, H. 2001, \aap, 369, 939

\bibitem[{{Williams} {et~al.}(2014){Williams}, {Peterson}, {Murphy}, {Gilbert},
  {Dalcanton}, {Dolphin}, \& {Jennings}}]{Williams+2014}
{Williams}, B.~F., {Peterson}, S., {Murphy}, J., {et~al.} 2014, \apj, 791, 105

\bibitem[{{Woosley} \& {Heger}(2006)}]{Woosley+2006}
{Woosley}, S.~E. \& {Heger}, A. 2006, \apj, 637, 914

\bibitem[{{Woosley} \& {Heger}(2007)}]{Woosley+2007a}
{Woosley}, S.~E. \& {Heger}, A. 2007, \physrep, 442, 269

\bibitem[{{Woosley} {et~al.}(2002){Woosley}, {Heger}, \&
  {Weaver}}]{Woosley+2002}
{Woosley}, S.~E., {Heger}, A., \& {Weaver}, T.~A. 2002, Reviews of Modern
  Physics, 74, 1015

\bibitem[{{Xiao} \& {Eldridge}(2015)}]{Xiao+2015}
{Xiao}, L. \& {Eldridge}, J.~J. 2015, \mnras, 452, 2597

\bibitem[{{Xiao} {et~al.}(2018){Xiao}, {Stanway}, \& {Eldridge}}]{Xiao+2018}
{Xiao}, L., {Stanway}, E.~R., \& {Eldridge}, J.~J. 2018, \mnras, 477, 904

\bibitem[{{Yaron} {et~al.}(2017){Yaron}, {Perley}, {Gal-Yam}, {Groh}, {Horesh},
  {Ofek}, {Kulkarni}, {Sollerman}, {Fransson}, {Rubin}, {Szabo}, {Sapir},
  {Taddia}, {Cenko}, {Valenti}, {Arcavi}, {Howell}, {Kasliwal}, {Vreeswijk},
  {Khazov}, {Fox}, {Cao}, {Gnat}, {Kelly}, {Nugent}, {Filippenko}, {Laher},
  {Wozniak}, {Lee}, {Rebbapragada}, {Maguire}, {Sullivan}, \&
  {Soumagnac}}]{Yaron+2017}
{Yaron}, O., {Perley}, D.~A., {Gal-Yam}, A., {et~al.} 2017, Nature Physics, 13,
  510

\bibitem[{{Yoon} {et~al.}(2017){Yoon}, {Dessart}, \& {Clocchiatti}}]{Yoon+2017}
{Yoon}, S.-C., {Dessart}, L., \& {Clocchiatti}, A. 2017, \apj, 840, 10

\bibitem[{{Yoon} {et~al.}(2012){Yoon}, {Gr{\"a}fener}, {Vink}, {Kozyreva}, \&
  {Izzard}}]{Yoon+2012a}
{Yoon}, S.-C., {Gr{\"a}fener}, G., {Vink}, J.~S., {Kozyreva}, A., \& {Izzard},
  R.~G. 2012, \aap, 544, L11

\bibitem[{{Yoon} {et~al.}(2010){Yoon}, {Woosley}, \& {Langer}}]{Yoon+2010}
{Yoon}, S.-C., {Woosley}, S.~E., \& {Langer}, N. 2010, \apj, 725, 940

\bibitem[{{Zahn}(1977)}]{Zahn1977}
{Zahn}, J.-P. 1977, \aap, 57, 383

\bibitem[{{Zapartas} {et~al.}(2017{\natexlab{a}}){Zapartas}, {de Mink},
  {Izzard}, {Yoon}, {Badenes}, {G{\"o}tberg}, {de Koter}, {Neijssel}, {Renzo},
  {Schootemeijer}, \& {Shrotriya}}]{Zapartas+2017}
{Zapartas}, E., {de Mink}, S.~E., {Izzard}, R.~G., {et~al.} 2017{\natexlab{a}},
  \aap, 601, A29

\bibitem[{{Zapartas} {et~al.}(2017{\natexlab{b}}){Zapartas}, {de Mink}, {Van
  Dyk}, {Fox}, {Smith}, {Bostroem}, {de Koter}, {Filippenko}, {Izzard},
  {Kelly}, {Neijssel}, {Renzo}, \& {Ryder}}]{Zapartas+2017a}
{Zapartas}, E., {de Mink}, S.~E., {Van Dyk}, S.~D., {et~al.}
  2017{\natexlab{b}}, \apj, 842, 125

\bibitem[{{Zwicky}(1957)}]{Zwicky1957}
{Zwicky}, F. 1957, {Morphological astronomy}

\end{thebibliography}

\end{document}